\documentclass[%
reprint,
amsmath,amssymb,
aps,
]{revtex4-2}

\usepackage{graphicx}
\usepackage[colorlinks=true,linkcolor=blue,anchorcolor=blue, citecolor=blue,urlcolor=blue]{hyperref}
\usepackage{array,color}

\begin{document}

\preprint{APS/123-QED}

\title{Fragile topology in nodal-line semimetal superconductors}

\author{Xiaoming Wang$^{1}$}
\author{Tao Zhou$^{1,2}$}%
\email{Corresponding author: tzhou@scnu.edu.cn}
\affiliation{$^1$Guangdong Provincial Key Laboratory of Quantum Engineering and Quantum Materials, School of Physics and Telecommunication Engineering, South China Normal University, Guangzhou 510006, China\\
	$^2$Guangdong-Hong Kong Joint Laboratory of Quantum Matter, Frontier Research Institute for Physics, South China Normal University, Guangzhou 510006, China
}

\begin{abstract}
	
	We study the band topology of the superconducting nodal-line semimetal (SC-NLSM) protected by the
	inversion symmetry with and without the spin-orbital coupling.
	Without the spin-orbital coupling, both the $s$-wave SC-NLSM and the chiral $p$-wave SC-NLSM
	are topologically nontrivial and can be described by the nonzero winding number. Based on the Wilson loop method, we verify that
	they are both
	the fragile topological superconductors, namely,
	their nontrivial band topologies can be moved off by coupling to additional topologically trivial bands. The fragile topological phase persists in spinful system with the time-reversal symmetry when a spin-orbital coupling term is added.
	For the spinful system, both the $p$-wave SC-NLSM and the $s$-wave SC-NLSM are second-order fragile topological superconductors.
	We propose that the fragile topology in the SC-NLSM system depends strongly on the degeneracy of the Majorana zero modes and the parity of the superconducting gap function.
	Interestingly, in presence of a vortex line,
	the spinful $s$-wave SC-NLSM system
	hosts two pairs of stable Majorana zero modes in the vortex core.

\end{abstract}

\maketitle

\section{Introduction }
Topological states have attracted extensive research interest in the field of the condensed matter physics, ranging from the fully gapped topological insulator to the gapless topological semimetal~\cite{Qi_2011,Hasan_2010,Wan2011,Ahn_2018,Zhang_2018}.
In the system bulk, the valance band and the conduction band of a topological nodal-line semimetal (NLSM) cross to each other and are partially inverted, forming a nodal loop at the Fermi energy.
At the system surface it has the drumhead surface state~\cite{Yang_2018,Fang2015,Fang_2016}. Previously several materials have been proposed to be the NLSM, such as PbTaSe$_{2}$~\cite{Bian_2016}, TlTaSe$_{2}$~\cite{Bian_20162}, InNbS$_{2}$/InNbSe$_{2}$~\cite{Du_2017}, In$_{x}$TaSe$_{2}$~\cite{Li_2021,Li_2020}, CaP$_3$~\cite{Xu_2017}, and ZrSiSe~\cite{Hu_2016}. The unusual bulk and surface states may provide a useful platform to exhibit various interesting physical phenomenon, such as the surface Chern insulator~\cite{Chen2019}, the three dimensional quantum hall effect~\cite{Molina_2018}, the
second-order topological insulator~\cite{Wang2019,Schindler_2018a,Schindler_2018b,Ahn_2020}, the $Z_{2}$ monopole~\cite{Ahn_2018,Fang2015,Wang2019}, and the possible superconducting state~\cite{Li_2020,Li_2021,Muechler_2019,Cheng_2020,Gao_2020,Aggarwal_2019,Zhang_2016,Setty_2017,PhysRevB.93.020506,Wang2017,Xu_2019,Chen_2019,Schnyder_2015,Shapourian2018}.

The concept of the fragile topology was proposed very recently~\cite{Po_2018,Wang2019,Ahn_2019,Bradlyn_2019,Hwang_2019,Alexandradinata_2020,Cano_2018,Song_2020a,Song_2020b,Wieder2018,Bouhon_2019}.
The Wanniner obstruction of a fragile topological material can be removed by coupling an additional topologically trivial band~\cite{Po_2018}. Generally, the fragile topological state do not host robust gapless edge state, thereby it may break the bulk and edge correspondence. Several materials, such as the twisted graphene~\cite{Lian_2020}, the Dirac semimetal with the higher order fermi arc~\cite{Wieder_2020}, and the photonic crystal~\cite{Alexandradinata_2020}, are proposed to exhibit fragile topological properties. An NLSM system with two nodal loops is also proposed to exhibit fragile topology~\cite{Po_2018,Wang2019}.

The method of the symmetry indicator has been widely used in searching for various topologically nontrivial systems~\cite{Slager2013,PhysRevX.7.041069,Zhang_2019,Vergniory_2019,Bradlyn_2017,Po_2017,Tang_2019}.
Recently, this method has been developed to search for the topological superconductor~\cite{Huang_2021,Skurativska_2020,Ono_2019,Sumita_2019,Geier_2020,Ono_2020}.
For the topological superconducting system, based on the symmetry indicator method, the band topology is described through the pairing symmetry and the representation of the filling bands in the normal state.
Moreover, in Ref.~\cite{Ono_2020}, the concept of the fragile topology was extended to the superconducting system. The topology of the fragile topological superconductor can also be diagnosed by the symmetry indicator method.
Similar to the fragile topological insulating system,
if a topologically nontrivial superconducting system becomes a topologically trivial superconductor when coupling to one or several topologically trivial bands, then this system is a fragile topological superconductor.
It was proposed that the system with two copies of Kitaev chains in presence of the inversion symmetry is a fragile topological~\cite{Ono_2020}. Searching for other fragile topological superconducting systems is of interest. Especially, as far as we know, the possible fragile topological superconducting state in the three dimensional system has not been
explored yet.

The Majorana Krammer pairs protected by the mirror symmetry and the rotational symmetry have been studied intensively~\cite{Ueno_2013,Kobayashi_2020,Zhang_2013,Chiu_2014}. However, as proposed in Ref.~\cite{Wang2019}, if the energy bands of the material invert twice or more, the edge states may be not stable. For the superconducting NLSM (SC-NLSM) system, the energy bands naturally invert twice due to the particle-hole symmetry. As a result, the edge states and the Majorana Krammer pairs may be not stable. Thus the SC-NLSM system may be a candidate to realize the fragile topology, with the topological properties being rather different from the stable topological system.

In this paper, we study the fragile topological properties of the SC-NLSM system with and without the spin orbital coupling. In the normal state, the NLSM system we considered has the inversion symmetry, the time reversal symmetry and the mirror symmetry. In the superconducting state,
previously the preferred pairing symmetry for an SC-NLSM system has not been identified yet.
The $s$-wave pairing symmetry was proposed by a recent experiment~\cite{Muechler_2019}, but the spin polarized surface state and the nontrivial (pseudo)-spin texture on the torus Fermi surface of the NLSM favor the spin triplet pairing instability~\cite{Wang2017,Shapourian2018}. So both pairing symmetries are considered in our present work. The fragile topology is verified based on the Wilson loop method. In the past the vortex line has been widely used to probe the topological properties~\cite{Yan_2020,Ghorashi_2020,Hosur_2011,Kheirkhah2020,Qin_2019}. Here a vortex line is introduced to both the $s$-wave and the $p$-wave SC-NLSM. The possible vortex states are studied numerically through the zero energy local density of states (LDOS). Our results indicate that the numerical results in presence of a vortex line may depend strongly on the specific systems and no general conclusions can be made for the vortex states in a fragile topological superconductor.

The structure of our paper is organized as follows. In Sec. \uppercase\expandafter{\romannumeral2}, we discuss the topological properties of the spinless SC-NLSM system. In Sec. \uppercase\expandafter{\romannumeral3}, we discuss the topological properties of the spinful SC-NLSM system in presence of the spin-orbital coupling. At last, we present a brief summary in Sec. \uppercase\expandafter{\romannumeral4}.

\section{The SC-NLSM system without the spin orbital coupling }
\subsection{Model and formalism }
We start with a model in the three dimensional system including the normal state term ($H_{N}$) and the superconducting pairing term ($H^{s/p}_{SC}$),
\begin{equation}
	H^{s/p}=H_{N}+H^{s/p}_{SC},
\end{equation}

\begin{equation}
	H_{N}=\sum_{\bf k}\left[
	\begin{split}
		\begin{array}{cc}
			H({\bf k})_{\mathrm{NLSM}}&0\\
			0&-H(-{\bf k})_{\mathrm{NLSM}}
		\end{array}
	\end{split}
	\right],
\end{equation}
\begin{equation}
	H^{s/p}_{SC}=\sum_{\bf k}\left[
	\begin{split}
		\begin{array}{cc}
			0&\Delta_{s/p}({\bf k})\\
			\Delta_{s/p}^{\dagger}({\bf k})&0
		\end{array}
	\end{split}
	\right].
\end{equation}
$H({\bf k})_{\mathrm{NLSM}}$ is a two-band model for the spinless (or preserving $SU(2)$ spin rotational symmetry) NLSM system~\cite{Fang2015}, with,
\begin{eqnarray}
	H({\bf k})_{\mathrm{NLSM}}&&=M({\bf k})\sigma_{x}+\lambda_{z}({\bf k})\sigma_{z}-\mu\sigma_0,
\end{eqnarray}
where $M({\bf k})=m-(t_{x}k_{x}^{2}+t_{y}k_{y}^{2}+t_{z}k_{z}^{2})$ and $\lambda_{z}({\bf k})=2\beta k_{z}$. \\

We consider two kinds of pairing symmetries, namely, the $s$-wave pairing symmetry and the chiral $p$-wave pairing symmetry.
The $s$-wave superconducting pairing term is expressed as,
\begin{eqnarray}
	H^{s}_{SC}=\sum_{{\bf k}}2\Delta_{s}\tau_{y}\otimes\sigma_{y},
\end{eqnarray}
and the chiral $p$-wave one is expressed as,\\
\begin{eqnarray}
	H_{SC}^{p}=\sum_{{\bf k}}2\Delta_{p}(k_{x}\tau_{x}\otimes\sigma_{0}+k_{y}\tau_{y}\otimes\sigma_{0}).
\end{eqnarray}
$\sigma_{i}$ and $\tau_{i}$ are the pauli matrices on the orbital channel and the particle-hole channel, respectively. $\Delta_{s/p}$ are the pairing amplitudes. $\mu$ represents the chemical potential. $t_{i}$ $(i=x,y,z)$ and $\beta$ are the hopping constants. The vector basis of Eq.(1) is $\Psi=[c_{a}({\bf k}),c_{b}({\bf k}),c_{a}^{\dagger}({\bf -k}),c_{b}^{\dagger}({\bf -k})]^{T}$, with the indices $a$ and $b$ representing two orbitals.

In the normal state, the nodal ring at the Fermi energy is protected by the the inversion symmetry ($I$), the time reversal symmetry ($T$), and the mirror symmetry ($M_{xy}$). The operators are expressed as: $T=\sigma_{x}K$ ($K$ is the complex conjugate operator), $I=M_{xy}=\sigma_{x}$, with $TH({\bf k})_{\mathrm{NLSM}}T^{-1}=H({\bf -k})_{\mathrm{NLSM}}$, $IH({\bf k})_{\mathrm{NLSM}}I^{-1}=H({\bf -k})_{\mathrm{NLSM}}$, and $M_{xy}H(k_{x},k_{y},k_{z})_{\mathrm{NLSM}}M_{xy}^{-1}=H(k_{x},k_{y},-k_{z})_{\mathrm{NLSM}}$.

To study the symmetry in the superconducting state, we first study the parity of the pairing term. From Eqs.(5) and (6), the momentum dependent pairing functions are rewritten as: $\Delta_{s}({\bf k})=\Delta_{s}i\sigma_{y}$, $\Delta_{p}({\bf k})=\Delta_{p}(k_{x}\sigma_{0}+k_{y}i\sigma_{0})$. The parities of superconducting pairing term are expressed as,
\begin{eqnarray}
	I\Delta_{s/p}({\bf k})I^{-1}&&=-\Delta_{s/p}({\bf -k}) \\
	M_{xy}\Delta_{s}({\bf k})M_{xy}^{-1}&&=-\Delta_{s}({\bf k}) \\
	T\Delta_{s}({\bf k})T^{-1}&&=-\Delta_{s}({\bf -k}) \\
	M_{xy}\Delta_{p}({\bf k})M_{xy}^{-1}&&=\Delta_{p}({\bf k})
\end{eqnarray}

The $s$-wave pairing term does not break any symmetry of the NLSM. Moreover, it induces some additional symmetries. The whole Hamiltonian with the $s$-wave pairing term is invariant under the particle-hole symmetry ($P^{s}$), the time reversal symmetry ($T^{s}$), the chiral symmetry ($S$), the inversion symmetry ($I^{s}$) and the mirror symmetry ($M^{s}_{xy}$). Based on the parity in Eq.(7)-(10), the symmetries of the whole $s$-wave SC-NLSM Hamiltonian are expressed as: $P^{s}H^{s}({\bf k})P^{-1}=-H^{s}(-{\bf k})$, $T^{s}H^{s}({\bf k})(T^{s})^{-1}=H^{s}(-{\bf k})$, $SH^{s}({\bf k})S^{-1}=-H^{s}({\bf k})$, $I^{s}H^{s}({\bf k})(I^{s})^{-1}=H^{s}(-{\bf k})$, and $M^{s}_{xy}H^{s}(k_{x},k_{y},k_{z})(M^{s}_{xy})^{-1}=H^{s}(k_{x},k_{y},-k_{z})$, with $P^{s}=\tau_{x}\otimes\sigma_{0}K$, $T^{s}=\tau_z\otimes\sigma_{x}K$, $S=T^{s}P^{s}=i\tau_y\otimes \sigma_x$, $I^{s}=\tau_z\otimes\sigma_x$, $M^{s}_{xy}=\tau_z\otimes\sigma_x$. Based on the ten fold Atland-Zirnbauer classification~\cite{Kitaev_2009,Ryu_2010}, it belongs to the class BDI. As for the  chiral $p$-wave pairing, the time reversal symmetry is naturally broken. The system is protected by the particle-hole symmetry ($P^{p}=\tau_x\otimes\sigma_{0}K$), the mirror symmetry ($M^{p}_{xy}=\tau_0\otimes\sigma_x$) and the inversion symmetry ($I^{p}=\tau_z\otimes\sigma_{x}$). Therefore, the topology of a  chiral $p$-wave SC-NLSM system is categorized to the class D.

In the lattice system and performing the partial Fourier transformation along the $z$ direction, the Hamiltonian is reexpressed as,
\begin{equation}
	\begin{split}
		H_{\mathrm{NLSM}}=&\sum_{{\bf k},z} \left[ (m-6) + 2t_{x}\cos k_{x} + 2t_{y}\cos k_{y} \right] [c^{\dagger}_{z,a}({\bf k})c_{z,b}({\bf k}) \\
		&+ c^{\dagger}_{z,b}({\bf k})c_{z,a}({\bf k})]+\sum_{{\bf k},z} [t_{z} c^{\dagger}_{z,a}({\bf k})c_{z+1,b}({\bf k})\\
		&+ t_z c^{\dagger}_{z,b}({\bf k})c_{z+1,a}({\bf k}) + h.c. ]  \\
		& +\sum_{z} \left[-i\beta_z  c^{\dagger}_{z,a}({\bf k})c_{z+1,a}({\bf k}) +i\beta_z c^{\dagger}_{z,b}({\bf k})c_{z+1,b}({\bf k}) \right.  \\
		&\left. + h.c.\right]-\mu\left[ c^{\dagger}_{z,a}({\bf k})c_{z,a}({\bf k})+c^{\dagger}_{z,b }({\bf k})c_{z,b}({\bf k}) \right].
	\end{split}
\end{equation}
The superconducting pairing term is then expressed as,
\begin{eqnarray}
	H^{s}_{SC}=\sum_{{\bf k},z}\left[ 2\Delta_{s}c^{\dagger}_{z,a}({\bf k})c^{\dagger}_{z,b}(-{\bf k})+h.c.\right],
\end{eqnarray}
and
\begin{eqnarray}
	H_{SC}^{p}=\sum_{{\bf k},z,\sigma}&&\left[ 2\Delta_{p}\left( \sin(k_{x})+i\sin(k_{y}) \right)c^{\dagger}_{z\sigma}({\bf k})c^{\dagger}_{z\sigma}(-{\bf k})\right.  \nonumber \\
	&&\left. +h.c. \right].
\end{eqnarray}
The whole Hamiltonian in the superconducting state [Eqs.(4-6)] can be rewritten as a $4N_z\times 4N_z$ matrix ($H=\Psi^{\dagger}M\Psi$) with the basis $\Psi=\left( c_{z,a},c_{z,b},c^{\dagger}_{z,a},c^{\dagger}_{z,b} \right)^{T} $ $(z=1,2,\cdots,N_z)$. $N_z$ is the number of sites along the $z$-direction.
The Green's function can be obtained through the matrix $M$. The matrix elements for the Green's function are expressed as below
\begin{eqnarray}
	G_{0}({\bf k},\omega)_{ij}=
	\sum_{n}\dfrac{u_{in}({\bf k}) u^{\dagger}_{jn}({\bf k})}{\omega-E_{n}({\bf k})+i\Gamma},
\end{eqnarray}
where $u_{in}({\bf k})$ and $E_{n}({\bf k})$ are the eigenvectors and the eigenvalues obtained by diagonalizing the matrix $M$.

The spatial dependent spectral function and the LDOS can then be expressed as,\\
\begin{eqnarray}
	A_{z}({\bf k},\omega)=-\frac{1}{\pi}\sum^2_{p=1}\mathrm{Im}G_{0}({\bf k},\omega)_{m+p,m+p},
\end{eqnarray}
and
\begin{eqnarray}
	\rho_{z}(\omega)=\frac{1}{N_{xy}}\sum_{\bf k}A_{z}({\bf k},\omega),
\end{eqnarray}
with $m=4(z-1)$. $N_{xy}$ is the number of sites of the $x-y$ plane.

We would also like to study the vortex bound states of the superconducting system~\cite{Yan_2020}. The vortex line is induced to the system through considering $\Delta({\bf r})=\Delta_{0}\tanh(\sqrt{x^{2}+y^{2}}/\zeta)e^{i\theta}$, with $\theta=\arctan(\dfrac{y}{x})$.  $\zeta$ is the coherence distance. The parameters in the following presented results are set as $\beta=t_{x}=t_{y}=t_{z}=1$, $m=3$, $\Delta_{s}=0.125$, $\Delta_{p}=0.35$, and $\zeta=4$.

\subsection{Fragile topology of the spinless SC-NLSM system}

\begin{figure}
	\includegraphics[scale=0.35]{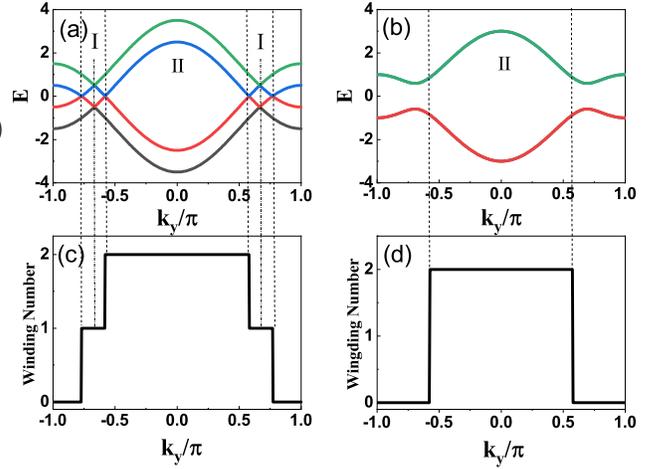}
	\caption{\label{fig:epsart} The energy bands as a function of $k_y$ with $(k_x,k_z)=(0,0)$ and the winding numbers of the $s$-wave SC-NLSM system and the $p$-wave SC-NLSM system with $k_x=0$ and $\mu=0$. 
		(a) The band structure of the $s$-wave SC-NLSM system. (b) The band Structure of the $p$-wave SC-NLSM system (two fold degeneracy). (c) Sum of the winding number of the occupied bands of the $s$-wave SC-NLSM system. (d) Sum of the winding number of the occupied bands of the $p$-wave SC-NLSM system. We have numerically confirm that the above results are robust for different chemical potentials with $0\leq \mu \leq 0.5$. \uppercase\expandafter{\romannumeral1} and \uppercase\expandafter{\romannumeral2} in panels (a) and (b) represent two different topologically nontrivial regions with winding numbers $1$ and $2$, respectively.}
	\label{fig:1}
\end{figure}

We now study the topological properties in the superconducting state with different pairing symmetries.
The band structures obtained through diagonalizing the Hamiltonian in the momentum space [Eq.(1)] with the $s$-wave pairing symmetry and the chiral $p$-wave pairing symmetry are presented in Figs.~\ref{fig:1}(a) and \ref{fig:1}(b), respectively. The corresponding winding numbers are plotted in Figs.~\ref{fig:1}(c) and \ref{fig:1}(d), respectively~\cite{supp}.

For the $s$-wave pairing symmetry, as is seen in Fig.~\ref{fig:1}(a), when $k_y$ increases from $-\pi$ to 0, the valance bands and the conduction bands cross twice at the Fermi energy. There are two topologically nontrivial regions indicated as \uppercase\expandafter{\romannumeral1} and \uppercase\expandafter{\romannumeral2} in Fig.~\ref{fig:1}(a). As presented in Figs.~~\ref{fig:1}(a) and \ref{fig:1}(c),
the energy band inverts once and twice at these two regions and the corresponding winding numbers are $1$ and $2$.
For the chiral $p$-wave pairing state, as is shown in Figs.~\ref{fig:1}(b) and \ref{fig:1}(d), the energy bands are fully gapped. They are two-fold degenerate when at the $k_z=0$ plane. The winding number is changed directly from $0$ to $2$ when the momentum crosses the normal state nodal ring.

\begin{figure}
	\includegraphics[width=8.5cm,height=9cm]{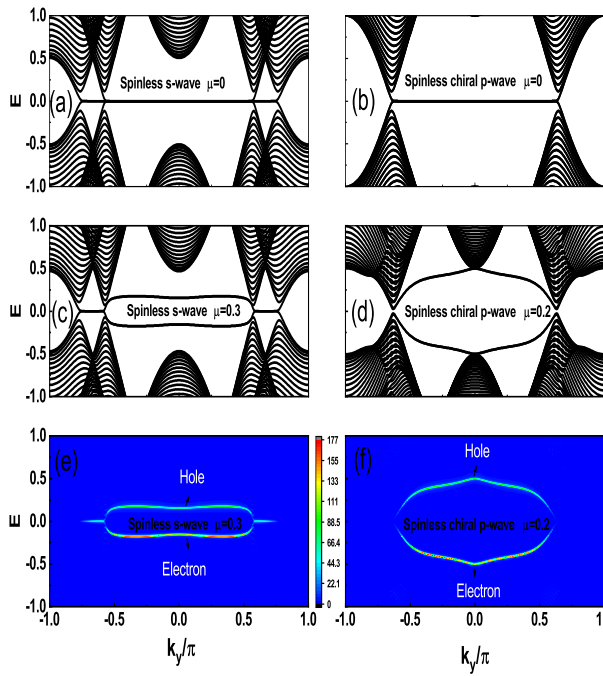}
	\caption{The energy bands and spectral functions considering the open boundary condition along the $z$-direction with $k_x=0$. (a-d) The energy bands with different pairing symmetries and different chemical potentials. Panels (e) and (f) are the intensity plots of the spectral functions at the system surface.
	}
	\label{fig:2}
\end{figure}

We turn to study the edge states through considering the open boundary condition along the $z$ direction and the periodic boundary condition along the $x-y$ plane [Eqs.(11-13)]. The corresponding energy bands for the $s$-wave pairing and the chiral $p$-wave pairing with different pairing symmetries and different chemical potentials are plotted in Figs.~\ref{fig:2}(a)-\ref{fig:2}(d), respectively. The spectral functions at the system surface ($z=1$) with a nonzero chemical potential are plotted in Figs.~\ref{fig:2}(e) and \ref{fig:2}(f). In presence of an additional chemical potential term, the previous four-fold degenerate edge states shown in Figs.\ref{fig:2}(a) and \ref{fig:2}(b) split into two two-fold degenerate edge states corresponding to the electron states and the hole states.
Generally, the chiral symmetry would be broken by the chemical potential term in the two band NLSM Hamiltonian, then the zero energy surface state shifts to the finite energy. However, in the $s$-wave superconducting state, the chiral symmetry is preserved even when the chemical potential is nonzero because of the particle-hole symmetry. As is shown in Fig.~\ref{fig:2}(c), the gap of the surface state is partially opened [at the region \uppercase\expandafter{\romannumeral2} of Fig.~\ref{fig:1}(a)], even though without breaking any symmetry. For the chiral $p$-wave pairing symmetry, as is shown in Fig.~\ref{fig:2}(d), the surface state is fully gapped in presence of an additional chemical potential term.

Our results indicate that when the winding number equals to 2,
the gapless edge states are not stable.
In principle, the nonzero winding number may generate Majorana zero modes at the system edges. However, the Majorana zero modes with equal spin at the same site are usually not stable. They will annihilate into the finite energy quasiparticles with
the particle-hole symmetry. As a result, an energy gap will be opened at the region \uppercase\expandafter{\romannumeral2} when an additional chemical potential term is added. This also implies that the system
may belong to a fragile topological superconductor.

\begin{figure}
	\includegraphics[scale=0.25]{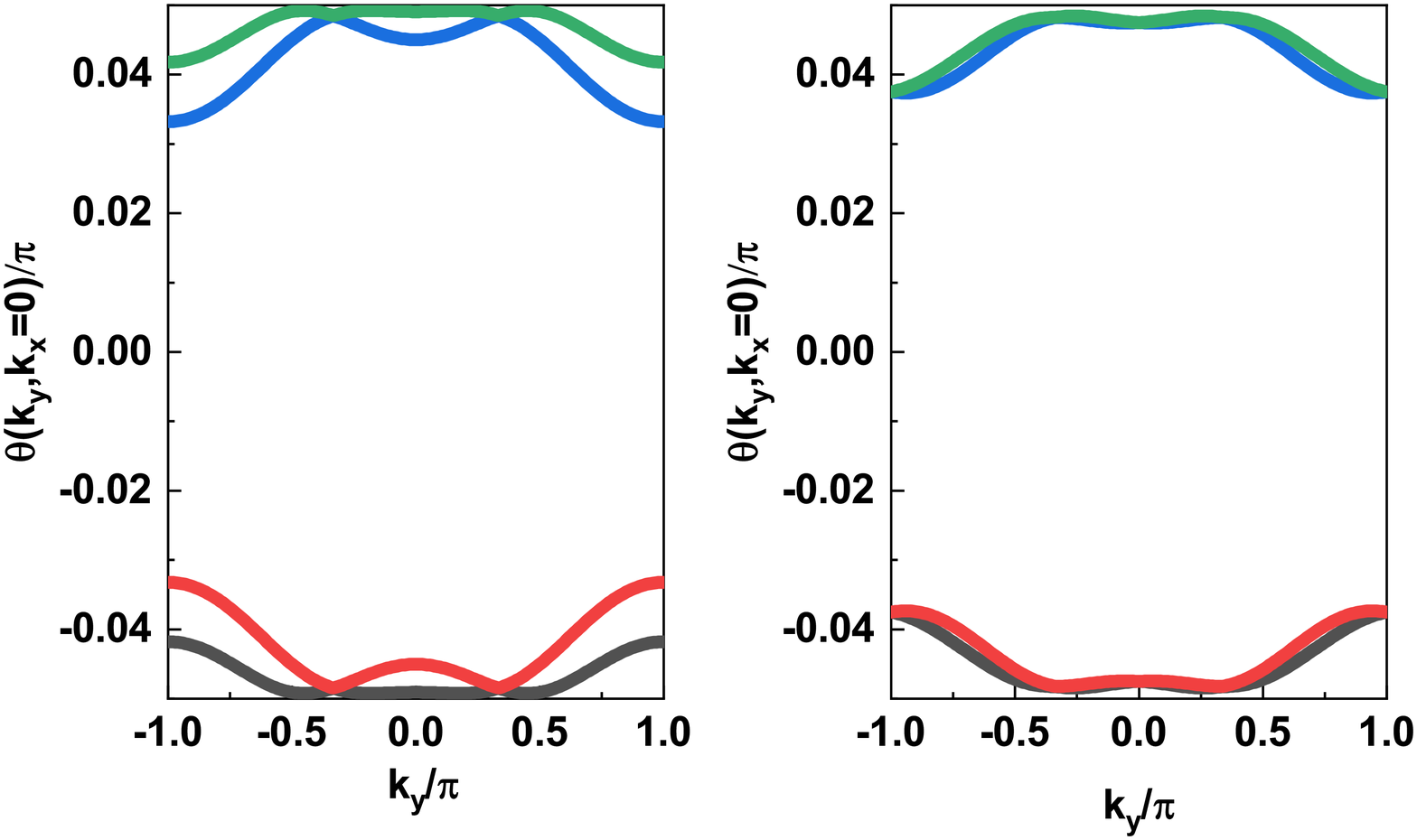}
	\caption{\label{fig:epsart} Numerical results of the Wilson loop spectra with the loop being along the $k_{z}$ direction after coupling four additional atomic orbitals [Eq.(17)].
		(a) The Wilson loop spectrum for the spinless $s$-wave system. (b) The Wilson loop spectrum for the chiral $p$-wave system.}
	\label{fig:3}
\end{figure}

As defined in Ref.~\cite{Ono_2020}, the Wannier obstruction of a fragile topological superconductor can be removed through coupling additional topologically trivial bands. Here we consider four additional particle-hole symmetric atomic orbitals and couple them into the original Hamiltonian of the SC-NLSM to verify its fragile topological properties. The whole Hamiltonian is expressed as,
\begin{eqnarray}
	H^{f}({\bf k})&&=H^{s/p}({\bf k})\oplus\left[
	\begin{split}
		\begin{array}{cc}
			-\sigma_{0}&0\\
			0&\sigma_{0}
		\end{array}
	\end{split}
	\right] \nonumber \\
	&&+3\left[
	\begin{split}
		\begin{array}{cc}
			0&\tau_{y}\otimes\sigma_{y}\\
			\tau_{y}\otimes\sigma_{y}&0
		\end{array}
	\end{split}\right],
\end{eqnarray}
where the base vector is expressed as $\Psi=[c_{a}({\bf k}),c_{b}({\bf k}),c_{a}^{\dagger}({\bf -k}),c_{b}^{\dagger}({\bf -k}),c_{1}({\bf k}),c_{2}({\bf k}),c_{1}^{\dagger}({\bf -k}),c_{2}^{\dagger}({\bf -k})]^{T}$. The inversion and particle-hole operators are rewritten as $I=(\tau_{z}\otimes \sigma_{x})\oplus (\tau_{z}\otimes \sigma_{x})$, $P=(\tau_{x}\otimes \sigma_{0}K)\oplus (\tau_{x}\otimes \sigma_{0}K)$. These two symmetries preserve in the whole coupled Hamiltonian.

Based on the Wilson loop method~\cite{Alexandradinata_2016,Wieder_2018,Alexandradinata_2014,supp},
the corresponding Wilson loop spectra from the above coupled Hamiltonian are calculated and presented in Fig.~\ref{fig:3}. As is seen, for both pairing symmetries, the Wilson loop spectra have been unwound, indicating that the systems become topologically trivial. Our numerical results verify directly the fragile topology of the present SC-NLSM system.

\subsection{Vortex bound states}

\begin{figure*}
	\includegraphics[scale=0.25]{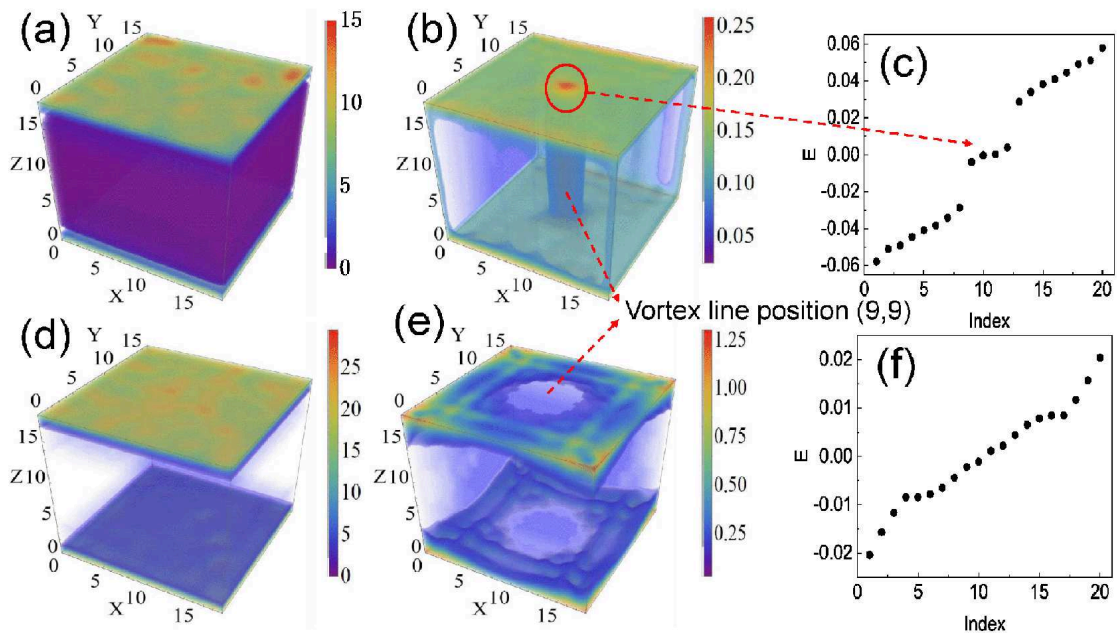}
	\caption{The LDOS at the zero energy with a vortex line along the $z$ direction for the $s$-wave and the $p$-wave pairing symmetries with the lattice size $19\times 19 \times 19$, with (a) $\mu=0$ for the chiral $p$-wave pairing; (b) $\mu=0.2$ for the chiral $p$-wave pairing; (d) $\mu=0$ for the $s$-wave pairing; (e) $\mu=0.3$ for the $s$-wave pairing. Panels (c) and (f) are the low energy eigenvalues of the Hamiltonian with (c) $\mu=0.2$ for the chiral $p$-wave pairing symmetry; (f) $\mu=0.3$ for the $s$-wave pairing symmetry, respectively.}
	\label{fig:4}
\end{figure*}

Let us study the possible vortex bound states in the spinless SC-NLSM system. The intensity plots of the zero energy LDOS spectra in presence of a vortex line for different chemical potentials and different pairing symmetries are displayed in Fig.~\ref{fig:4}.

We first discuss the possible vortex bound states in the chiral $p$-wave superconducting state. When the chemical potential is zero [Fig.~\ref{fig:4}(a)], the zero energy states exist at the system surface with $z=1$ and $z=N_z$. In this case there are no bound states existing. As the chemical potential increases to $\mu=0.2$, the vortex bound states emerge in the vortex core, as is seen in Fig.~\ref{fig:4}(b), which is similar to the surface phase transition proposed in the fully gapped second-order topological superconductor with a vortex line~\cite{Ghorashi_2020}.
On the other hand, here the energy of the bound state is not exactly at the zero energy, as is seen in Fig.~\ref{fig:4}(c). Our numerical results indicate that no Majorana modes exist in the vortex core.

Since the spinless SC-NLSM system has mirror symmetry, the Hamiltonian of the chiral $p$-wave SC-NLSM can be block diagonalized into two $2\times 2$ matrices with,
\begin{eqnarray}
	H^{p}({\bf k})&=H^{M_{xy}^{p}}_{-1}({\bf k})\oplus H^{M_{xy}^{p}}_{1}({\bf k}).
\end{eqnarray}
The subscripts $\pm1$ are the eigenvalues of the mirror symmetry operator ($M^{p}_{xy}$). The subsector Hamiltonian $H^{M_{xy}^{p}}_{\pm 1}$ at the $k_{z}=0$ plane is expressed as
\begin{eqnarray}
	H^{M_{xy}^{p}}_{-1}({\bf k})&&=-M({\bf k})\sigma_{z}+2\Delta_{p}k_{x}\sigma_{x}+2\Delta_{p}k_{y}\sigma_{y}, \\
	H^{M_{xy}^{p}}_{1}({\bf k})&&=-M({\bf k})\sigma_{z}+2\Delta_{p}k_{x}\sigma_{x}-2\Delta_{p}k_{y}\sigma_{y}.
\end{eqnarray}
The two subsector Hamitonians have a nontrivial Chern number ($1/-1$) and they both have the particle-hole symmetry $P$ with $P=\sigma_{x}K$. Based on the ten fold AZ classification, they both belong to the class D and can transform to each other via the spinless time reversal operation. In presence of a vortex line, in principle two pairs of Majorana zero modes should appear. However, similar to the case of the surface states, here the Majorana states are not stable. They will annihilate and four finite energy bound states with the particle-hole symmetry emerge, as presented in Fig.~\ref{fig:4}(c).

The above numerical results can be understood further through analysing the topological invariant. The topological invariant of the vortex line for the gapped $p$-wave system is defined as~\cite{supp,Fu_2007a,Kitaev_2001,Budich_2013},
\begin{eqnarray}
	{\mathrm v}_{p}={\mathrm{sgn}}\left\lbrace {\mathrm{Pf}}\left[ H^{p}_{M}(\pi) \right]  \right\rbrace {\mathrm{sgn}}\left\lbrace {\mathrm{Pf}} \left[ H^{p}_{M}(0) \right]  \right\rbrace. \nonumber \\
\end{eqnarray}
Here ${\mathrm v}_{p}=-1$/+1 describes the topologically nontrivial/trivial vortex line, with/without the Majorana bound states at the end of the vortex line. Based on our calculation~\cite{supp}, here ${\mathrm v}_{p}$ equals to $+1$, being independent on the chemical potential $\mu$. This indicates that the vortex bound states obtained in Fig.~\ref{fig:4}(b) are indeed not Majorana bound states, while on the other hand, we cannot conclude that here the vortex line is topologically trivial. It was proposed in Ref.~\cite{Ono_2020} that two copies of Kitaev chains protected by the inversion symmetry is a fragile topological superconductor and hosts fragile end states. Here the topological properties of the vortex line in a chiral $p$-wave SC-NLSM system still need further studies.

The numerical results for the case of the $s$-wave compound are presented in Figs.~\ref{fig:4}(c) and \ref{fig:4}(d). For the zero chemical potential, as is presented in Figs.~\ref{fig:4}(c), the result is qualitatively the same with that of the chiral $p$-wave compound, namely, the zero energy states appear at the system surfaces and no vortex bound states exist. Interestingly, when the chemical potential increases, as is seen in Fig.~\ref{fig:4}(d), the result is significantly different from that in the chiral $p$-wave system~\cite{note}. There are no bound states in the vortex core. Instead, the intensity of the zero energy LDOS is nearly zero in the vortex region. Outside of the vortex region, the zero energy LDOS increases and it reaches the maximum value at the system corners.

The above results for the $s$-wave compound can be understood through exploring the superconducting gap magnitudes. In the normal state, the flat bands at the system surface will shift to the energy $E=\mu$ when an additional chemical potential $\mu$ is added to the system. On the other hand, in the $s$-wave superconducting state, the flat bands still exist due to the particle-hole symmetry. The size of the flat band region will increase monotonously when the gap magnitude increases~\cite{supp}. As a result, the zero energy LDOS depends strongly on the superconducting order parameter magnitudes when the chemical potential is nonzero. Here in presence of a vortex, the gap magnitude is suppressed in the vortex core and is largest at the system corners, leading to the distribution of the zero energy LDOS presented in Fig.~\ref{fig:4}(d).

\begin{figure}
	\includegraphics[scale=0.3]{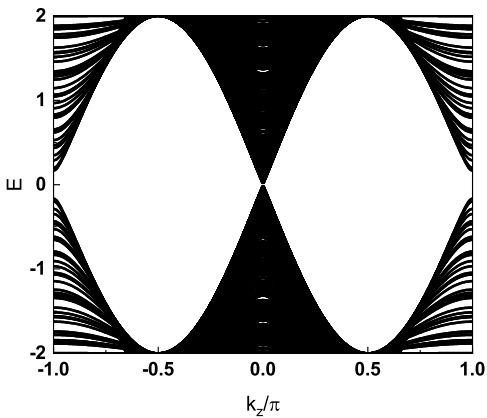}
	\caption{The band structure as a function of $k_{z}$ with a vortex line along the $z$ direction for the spinless $s$-wave SC-NLSM system with $\mu=0.3$.}
	\label{fig:5}
\end{figure}

Let us discuss the numerical results of the $s$-wave SC-NLSM system in more detail. In presence of a vortex line along the $z$ direction, the momentum along the $z$ direction ($k_z$) is still a good quantum number. Considering the open boundaries along the $x$ and $y$ directions with the system size $41\times 41$ and putting a vortex line at the site $(x,y)=(20,20)$, the band structure as a function of $k_z$ is presented in Fig.~\ref{fig:5}. As is seen, the system is indeed gapless with the energy bands crossing the Fermi energy at $k_z=0$. Actually, for a $s$-wave SC-NLSM, the gapless states are robust and protected by the mirror symmetry.

We block diagonalize the Hamiltonian for the $s$-wave SC-NLSM system based on the eigenvectors and eigenvalues of the mirror symmetry operator $M_{xy}^{s}$, with
\begin{eqnarray}
	H^{s}({\bf k})&&=H^{M_{xy}^{s}}_{-1}({\bf k})\oplus H^{M_{xy}^{s}}_{1}({\bf k}) \\
	H^{M_{xy}^{s}}_{-1}({\bf k})&&=-M({\bf k})\sigma_{0}-2\Delta_{s}\sigma_{x}-\mu\sigma_z \\
	H^{M_{xy}^{s}}_{1}({\bf k})&&=M({\bf k})\sigma_{0}-2\Delta_{s}\sigma_{x}+\mu\sigma_z.
\end{eqnarray}
$H^{M_{xy}^{s}}_{\pm1}$ are subsectors of the system.
Each subsector generates one nodal ring. Digonalizing the Hamiltonian matrix $H^{M_{xy}^{s}}_{-1}({\bf k})$/$H^{M_{xy}^{s}}_{1}({\bf k})$, we have $E_{-1,a/b}=M({\bf k})\pm\sqrt{4\Delta^{2}_{s}+\mu^{2}}$ and $E_{1,a/b}=-M({\bf k})\pm\sqrt{4\Delta^{2}_{s}+\mu^{2}}$, respectively. We have
\begin{eqnarray}
	E_{-1,a}&&=-E_{1,b} \\
	E_{-1,b}&&=-E_{1,a}.
\end{eqnarray}
For the whole $4\times4$ Hamiltonian, two nodal rings exist at the mirror plane ($k_{z}=0$), consistent with the numerical results of the band structure presented in Fig.~\ref{fig:1}.

When a vortex line along the $z$ direction is induced to the system and considering the open boundaries along the $x$ and $y$ directions, the Hamiltonian Eq.(1) can be rewritten as:
\begin{eqnarray}
	H^{s,vortex}&&=\sum_{{\bf k_{z}},y,z}M(k_{z},x,y)c_{a/b}^{\dagger}(k_{z},x,y)c_{a/b}(k_{z},x,y) \nonumber \\
	&&+e^{i\theta(x,y)}\Delta_{s}(k_{z},x,y)c_{a}^{\dagger}(k_{z},x,y)c_{b}^{\dagger}(-k_{z},x,y)+h.c. \nonumber \\
\end{eqnarray}
In the presence of vortex line, the mirror symmetry and the inversion symmetry preserve~\cite{supp}. At the mirror plane $k_{z}=0$, the Hamiltonian can be expressed as,
\begin{eqnarray}
	H^{s,vortex}({k_{z}=0})&&=H^{M_{xy}^{s,vortex}}_{-1}({k_{z}=0})\oplus H^{M_{xy}^{s,vortex}}_{1}({k_{z}=0}). \nonumber \\
\end{eqnarray}
Here the $H^{M_{xy}^{s,vortex}}_{\pm1}$ can be obtained through the Fourier transformation of $H^{M_{xy}^{s}}_{\pm1}({\bf k})$ and adding an additional vortex line into the system.
Thus the relation in Eqs. (25) and (26) should still satisfy. The energy bands in each subspace ($H^{M_{xy}^{s,vortex}}_{-1}({k_{z}=0})/H^{M_{xy}^{s,votex}}_{1}({k_{z}=0})$) will cross to each other and generate nodal points at $k_{z}=0$.

\section{The SC-NLSM system with the spin-orbital coupling}
\subsection{Model and formalism }
Generally the spin-orbital coupling (or the broken $SU(2)$ rotational symmetry) will directly change the topological classification of the system and lead to different topological properties. In a real material system, the spin-orbital coupling usually cannot be ignored. Here we start to discuss the topological properties of the spinful SC-NLSM system in the presence of the spin orbital coupling. The Hamiltonian is expressed as,
\begin{equation}
	H^{s/p}_{soc}({\bf k})=\sum_{\bf k}\left[
	\begin{split}
		\begin{array}{cc}
			h({\bf k})&0\\
			0&-h^{*}({\bf -k})
		\end{array}
	\end{split}
	\right] +\sum_{{\bf k}}h^{s/p}_{sc}({\bf k}).
\end{equation}
$h({\bf k})$ is the Hamiltonian of NLSM in presence of the spin-orbital coupling, with
\begin{eqnarray}
	h({\bf k})=&&M({\bf k}) s_{0}\otimes\sigma_{x}+\lambda_{z}({\bf k}) s_{0}\otimes\sigma_{z}-\mu s_{0}\otimes\sigma_{0} \nonumber \\
	&&+R\left[ \sin(k_{x}) s_{y}\otimes\sigma_{y}-\sin(k_{y}) s_{x}\otimes\sigma_{y}\right],
\end{eqnarray}
where $R$ is the Rashba spin orbital coupling strength.

$h^{s/p}_{soc}({\bf k})$ represents the superconducting pairing, with
\begin{eqnarray}
	h^{s}_{sc}({\bf k})=-2\Delta^{soc}_{s}\tau_{y}\otimes s_{y}\otimes\sigma_{z},
\end{eqnarray}
and
\begin{eqnarray}
	h^{p}_{sc}({\bf k})=2\Delta^{soc}_{p}\left[ \sin(k_{x})\tau_{x}\otimes  s_{0}\otimes\sigma_{0}-\sin(k_{y})\tau_{y}\otimes s_{z}\otimes\sigma_{0}\right]. \nonumber \\
\end{eqnarray}
$\sigma_{i}$, $s_{i}$, $\tau_{i}$ are pauli matrices and in the orbital, spin, and particle-hole channels, respectively. The base vector is expressed as $\Psi=\left[ c_{a,\uparrow}({\bf k}),c_{b,\uparrow}({\bf k}),c_{a,\downarrow}({\bf k}),c_{b,\downarrow}({\bf k}),c^{\dagger}_{a,\uparrow}({\bf -k}),
c^{\dagger}_{b,\uparrow}({\bf -k}),\right. \\
\left.c^{\dagger}_{a,\downarrow}({\bf -k}), c^{\dagger}_{b,\downarrow}({\bf -k}) \right]^{T} $. The parameters are set as $\beta=1$, $t_{x,y,z}=1$, $m=3$, $R=0.4$, $\Delta^{soc}_{s}=0.25$, $\Delta^{soc}_{p}=1$, $\mu=0$.

In the normal state with $\Delta^{soc}_{s/p}=0$, the above Hamiltonian is protected by the inversion symmetry ($g_{1}=s_{0}\otimes\sigma_{x}$), the time reversal symmetry ($g_{2}=is_{y}K\otimes\sigma_{x}$), and the mirror symmetry ($g_{z}=is_{z}\otimes\sigma_{x}$).

To study the symmetry of the system in the superconducting state, we rewrite the
$s$-wave and the $p$-wave gap functions as, $\Delta^{soc}_{s}({\bf k})=2\Delta^{soc}_{s}is_{y}\otimes\sigma_{z}$ and $\Delta^{soc}_{p}({\bf k})=2\Delta^{soc}_{p}\left[ \sin(k_{x})  s_{0}\otimes\sigma_{0}+i\sin(k_{y}) s_{z}\otimes\sigma_{0}\right]$.
The parities of the superconducting order parameters are expressed as,
\begin{eqnarray}
	g_1\Delta^{soc}_{s}({\bf k})g_1^{-1}&&=-\Delta^{soc}_{s}({\bf -k}) \\
	g_2\Delta^{soc}_{s}({\bf k})g_2^{-1}&&=-\Delta^{soc}_{s}({\bf -k}) \\
	g_1\Delta^{soc}_{p}({\bf k})g_1^{-1}&&=-\Delta^{soc}_{p}({\bf -k}) \\
	g_2\Delta^{soc}_{p}({\bf k})g_2^{-1}&&=-\Delta^{soc}_{p}({\bf -k}) \\
	g_{z}\Delta^{soc}_{p}({\bf k})g_{z}^{-1}&&=\Delta^{soc}_{p}(k_{x},k_{y},-k_{z}).
\end{eqnarray}
Based on the above results, The $s$-wave/$p$-wave SC-NLSM both belongs to the class DIII. For both  systems, the superconducting parameter is odd with the inversion symmetry operator. Note that in presence of the spin-orbital coupling, the energy bands are both fully gapped.

\begin{figure}
	\includegraphics[scale=0.34]{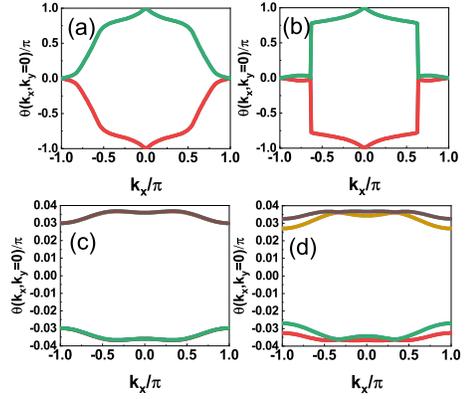}
	\caption{Wilson loop spectra of the spinful SC-NLSM system with the loop being defined along the $k_{z}$ direction. (a) Wilson loop spectrum of the spinful $s$-wave SC-NLSM system. (b) Wilson loop spectrum of the spinful $p$-wave SC-NLSM system. (c) Similar to the panel (a) but coupling additional topologically trivial bands. (d)
		Similar to panel (b) but coupling additional topologically trivial bands.	
	}
	\label{fig:6}
\end{figure}

\subsection{Fragile topology of the spinful SC-NLSM system}

Let us study the topological properties of the spinful SC-NLSM system based on the Wilson loop method~\cite{Alexandradinata_2016,Wieder_2018,Alexandradinata_2014,supp}. The numerical results of Wilson loop spectra for the $s$-wave and $p$-wave pairing symmetries with the winding direction along the $k_z$ direction are presented in Figs.~\ref{fig:6}(a) and ~\ref{fig:6}(b), respectively. As is seen, for both pairing symmetries, the Wilson loop spectra are winding, indicating that the spinful SC-NLSM system is indeed topologically nontrivial.

We now consider additional atomic bands coupling to the present spinful SC-NLSM system and make sure that the symmetry of the whole system does not change.
The whole Hamiltonian is expressed as,
\begin{eqnarray}
	H^{f^{'}}({\bf k})&&=H_{soc}^{s/p}({\bf k})_{soc}({\bf k})\oplus\left[
	\begin{split}
		\begin{array}{cc}
			-s_{0}\otimes \sigma_{0}&0\\
			0&s_{0}\otimes \sigma_{0}
		\end{array}
	\end{split}
	\right]\nonumber\\
	&&+3.5\left[
	\begin{split}
		\begin{array}{cc}
			0&\tau_{y}\otimes s_{y}\otimes\sigma_{y}\\
			\tau_{y}\otimes s_{y}\otimes\sigma_{y}&0
		\end{array}
	\end{split}
	\right].
\end{eqnarray}
The numerical results of the Wilson loop spectra from the above coupled Hamiltonian for the $s$-wave pairing symmetry and the $p$-wave pairing symmetry are presented in Figs.~\ref{fig:6}(c) and \ref{fig:6}(d). As is seen, the Wilson loop spectra are unwound when coupling additional atomic bands, indicating that the spinful SC-NLSM compounds should indeed be fragile superconductors.
Actually, previously the fragile topology is usually proposed in the spinless system. It is of rather interest to realize it in a spinful system. We have also checked numerically that when the winding direction is along the $k_x$ direction or the $k_y$ direction, the results are qualitatively the same with those along the $k_z$ direction.

\begin{figure}
	\includegraphics[scale=0.8]{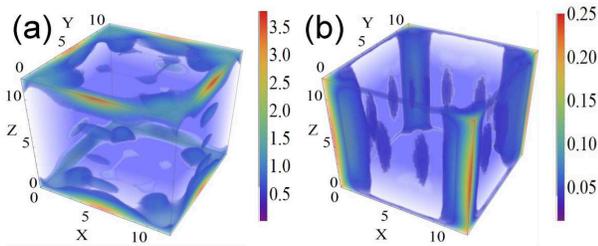}
	\caption{The intensity plots of the zero energy LDOS in the real space considering the open boundary condition with the lattice size $13\times 13 \times 13$ and $\mu=0$.
		(a) Zero energy LDOS for the spinful $p$-wave SC-NLSM system. (b) Zero energy LDOS for the spinful $s$-wave SC-NLSM system.
	}
	\label{fig:7}
\end{figure}

Considering the open boundaries, we present the zero energy LDOS with the $s$-wave and $p$-wave pairing symmetries in Fig.~\ref{fig:7}. As is seen, the zero energy states emerge at the system hinges for both two pairing symmetries we considered, indicating that these two systems may be second-order topological superconductors. 

For the $p$-wave system, the edge states emerge at the boundaries of the $z=1$ and $z=N_z$ planes. It seems  that these two planes can be seen as the two-dimensional Chern insulator, indicating that the topology of the $p$-wave SC-NLSM system may be described by the mirror Chern number.
The mirror symmetry operator $g_z$ has two eigenvalues $i$ and $-i$. The spinful $p$-wave SC-NLSM Hamiltonian can be block diagonalized as,
\begin{eqnarray}
	H^{p}_{soc}({\bf k})&&=H^{p,g_{z}}_{i}\oplus H^{p,g_{z}}_{-i}, \\
	H^{p,g_{z}}_{-i}&&=M({\bf k})\tau_{z}\otimes\sigma_{z}+R\sin(k_{x})\tau_{0}\otimes\sigma_{x} \nonumber \\
	&&+R\sin(k_{y})\tau_{z}\otimes\sigma_{y} +2\Delta^{soc}_{p}\sin(k_{x})\tau_{x}\otimes\sigma_{0} \nonumber \\
	&&+2\Delta^{soc}_{p}\sin(k_{y})\tau_{y}\otimes\sigma_{z}, \\
	H^{p,g_{z}}_{i}&&=-M({\bf k})\tau_{z}\otimes\sigma_{z}+R\sin(k_{x})\tau_{0}\otimes\sigma_{x} \nonumber \\
	&&+R\sin(k_{y})\tau_{z}\otimes\sigma_{y}+2\Delta^{soc}_{p}\sin(k_{x})\tau_{x}\otimes\sigma_{0} \nonumber \\
	&&+2\Delta^{soc}_{p}\sin(k_{y})\tau_{y}\otimes\sigma_{z}.
\end{eqnarray}
The Chern numbers of the Hamiltonian in the subsector of system ($H^{p,g_{z}}_{-i}$ and $H^{p,g_{z}}_{i}$) are $2$ and $-2$, respectively. Then the mirror Chern number should be $2$, consistent with the nontrivial topology of this system. Based on the bulk and edge correspondence, those nonzero Chern number ($2$ and $-2$) in the two subsectors should in principle generate two spin-up edge states and two spin-down ones. Because of the particle-hole symmetry, each subsector will generate a pair of spin polarized Majorana zero modes. Two spin polarized Majorana zero modes at the same site are not stable, leading to the fragile topology of this system.

\begin{figure}
	\includegraphics[scale=0.36]{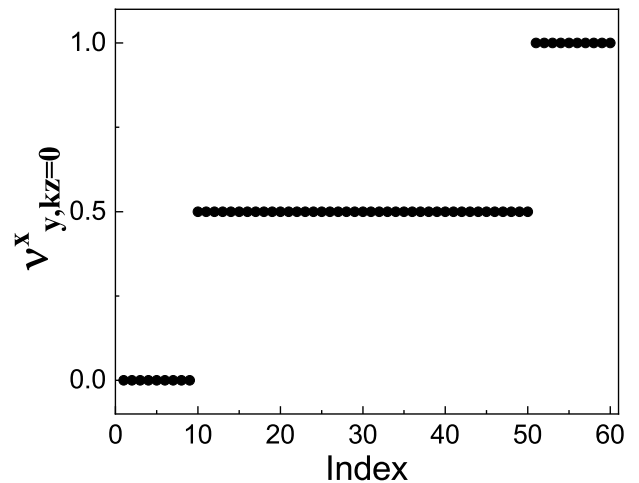}
	\caption{The wannier spectra as a function of the band index for
		the spinful $s$-wave SC-NLSM with the open boundary condition along the
		$x$-direction and periodic boundary condition along the $y-z$ plane~\cite{supp}.
	}
	\label{fig:8}
\end{figure}

For the $s$-wave pairing symmetry, the second-order topology can be investigated through the Wannier spectra~\cite{PhysRevB.104.134508,supp}. Considering the open boundary condition along the $x$-direction and periodic boundary condition along the $y-z$ plane, the Wannier spectra
with the winding direction along the $k_y$-direction is plotted in Fig.~\ref{fig:8}.
Our numerical results indicate that a quantized Wannier spectra with $v^{x}_{y,k_{z}=0}=0.5$ exist, corresponding to the hinge polarization. The Wannier spectra are highly degenerate at this value. Based on the argument provided in Ref.~\cite{PhysRevB.104.134508}, the degeneracy determines the number of majorana zero modes per hinge.
Therefore, here each hinge may host mutiple channels of Majorana zero modes, leading to the fragile topology of this system.

Usually, the fragile topological insulator is expected to be realized in the spinless system or
the topological crystalline insulator. Here we have verified that in
a superconducting spinful system with the time reversal symmetry, the fragile topology
can also be realized.
Our results indicate that the fragile topology depends strongly on the parity of the gap function under the inversion operation.
Here both pairing functions are odd under the inversion operation, as is seen in Eqs.(33) and (36).
We have also checked numerically the topological properties with other pairing symmetries.
For a chiral $p$-wave pairing symmetry, the pairing function is also odd under the inversion operation. The system is also a fragile topological superconducting system. For the $s$-wave pairing symmetry with $\Delta_{s}({\bf k})=\Delta_{s}is_{y}\otimes(\sigma_{x}/\sigma_{0})$, the gap function is even under the inversion operation. We have checked and confirmed that in this case the system is topologically trivial.
Besides the inversion symmetry, the fragile topology may also exist in other symmetry configurations.
The fragile topology of the SC-NLSM in a spinful system and the band topology with other possible symmetry configurations are of interest and worth further studies.

\subsection{Vortex bound states}

\begin{figure}
	\includegraphics[scale=0.76]{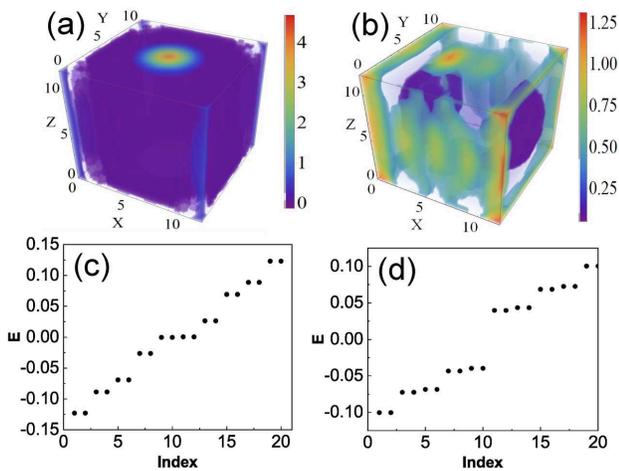}
	\caption{The LDOS at the zero energy for the spinful SC-NLSM system in presence of a vortex line and the corresponding low energy eigenvalues of the Hamiltonian with $\mu=0$. (a) The zero energy LDOS for the spinful $s$-wave SC-NLSM system.
		(b) The zero energy LDOS for the spinful $p$-wave SC-NLSM system.	(c) Low energy eigenvalues of the spinful $s$-wave SC-NLSM Hamiltonian. (d) Low energy eigenvalues of the spinful $p$-wave SC-NLSM Hamiltonian.
	}
	\label{fig:9}
\end{figure}

Let us turn to study the vortex bound state for the SC-NLSM system with the spin-orbital coupling.
The intensity plots of the zero energy LDOS in presence of a vortex line for the $s$-wave SC-NLSM system and the $p$-wave SC-NLSM system are presented in Figs.~\ref{fig:9}(a) and \ref{fig:9}(b), respectively. The corresponding low energy eigenvalues of the Hamiltonian are presented in Figs.~\ref{fig:9}(c) and \ref{fig:9}(d).
As is shown in Fig.~\ref{fig:9}(a), the bound states emerge in the vortex core for the $s$-wave SC-NLSM system. The energy of the bound states is exactly at the zero energy, as is seen in Fig.~\ref{fig:9}(c). Note that, we have checked numerically that the four zero energy eigenvalues are rather stable and does not depend on the chemical potential. On the other hand, for the $p$-wave SC-NLSM system, no obvious bound states are obtained. We have checked numerically that the results are qualitatively the same with $0\leq \mu\leq 2$.

In the normal state, the Hamiltonian  $h({\bf k})$ has the $C_{4}$ rotational symmetry. In the superconducting state, the $s/p$ pairing term breaks the $C_{4}$ rotational symmetry and only the $C_{2}$ rotational symmetry exists with $C_{2}=\exp(i\pi s_{z}/2)\otimes\sigma_{0}$. The pairing terms are odd under the $C_{2}$ rotational operation, with $C_{2}\Delta^{soc}_{s}({\bf k})/\Delta^{soc}_{p}({\bf k})(C_{2})^{-1}=-\Delta_{s}(-k_{x},-k_{y},k_{z})/\Delta_{p}(-k_{x},-k_{y},k_{z})$. As is discussed in Ref.~\cite{Ueno_2013}, the symmetry protected Majorana Krammer pairs in a topological superconductor should satisfy the condition that its superconducting gap is odd under certain symmetry operation. Here the four zero energy vortex bound states in the $s$-wave SC-NLSM system can be stabilized by the $C_{2}$ rotational symmetry, which is similar to the double Majorana vortex bound states studied in Refs.~\cite{Kobayashi_2020,hu2021competing}.
Note that, the $C_{2}$ operator has two eigenvalues, as a result, two pairs of Majorana vortex bound states can survive.
For the $p$-wave SC-NLSM system, the two subsectors $H^{p,g_{z}}_{i}$ and $H^{p,g_={z}}_{-i}$ in Eqs. (39-41) have a Chern number $2$ and $-2$, respectively. In presence of a vortex, in principle, each subsector should generate two pairs of majorana zero modes. Here the $C_{2}$ rotational symmetry cannot stabilize four pairs of Majorana zero modes. Thus it is understandable that for the $p$-wave SC-NLSM system, the Majorana zero modes cannot survive. In a fragile topological superconducting system with multiple topologically nontrivial bands, the vortex line topology is a rather complicated issue and still needs further studies.

\section{Summary}
In summary, we have studied the topological properties of the SC-NLSM system considering the $s$-wave pairing symmetry and the $p$-wave pairing symmetry. For the spinless system, we have verified that the system should be a fragile topological superconductor for both two pairing symmetries we considered. For the spinful system in presence of the spin-orbital coupling and with the time reversal symmetry, the system is a second-order fragile topological superconductor, with the fragile topology is protected by the inversion symmetry. Our results indicate that the fragile topology in the SC-NLSM system comes from the multiple deneracy of Majorana zero modes. When the
symmetry in a system cannot stablize these zero modes, the system may become topologically trival one or a fragile topological one. This criteria may be applied to search for other fragile topological superconductors.

In presence of a vortex line,
the possible vortex states in the SC-NLSM systems are discussed. For the spinless $p$-wave system, the vortex bound states exist at the vortex core while no zero energy states exist. For the spinless $s$-wave system, the energy bands are gapless even in presence of a vortex line and no vortex bound states are obtained. Here the zero modes may exist at the system surface, depending on the parameters, while for all of the parameters we considered, no zero modes are obtained in the vortex core. These results are rather different from those obtained in usual stable topological superconductors.
For the spinful system in presence of the spin-orbital coupling, the robust zero energy modes exist in the vortex core for the $s$-wave system, while no vortex bound states exist for the $p$-wave system. Therefore, in a fragile topological superconductor, whether the bound states and Majorana zero modes exist in the vortex core may depend strongly on the real system.
Our results may help to understand the fragile topology in the superconducting system further.

\begin{acknowledgments}
	This work was supported by the NSFC (Grant No. 12074130), the Natural Science Foundation of Guangdong Province (Grant No. 2021A1515012340), and Science and Technology Program of Guangzhou (Grant No. 2019050001 and No. 202102080434).
\end{acknowledgments}

\begin{thebibliography}{79}%
	\makeatletter
	\providecommand \@ifxundefined [1]{%
		\@ifx{#1\undefined}
	}%
	\providecommand \@ifnum [1]{%
		\ifnum #1\expandafter \@firstoftwo
		\else \expandafter \@secondoftwo
		\fi
	}%
	\providecommand \@ifx [1]{%
		\ifx #1\expandafter \@firstoftwo
		\else \expandafter \@secondoftwo
		\fi
	}%
	\providecommand \natexlab [1]{#1}%
	\providecommand \enquote  [1]{``#1''}%
	\providecommand \bibnamefont  [1]{#1}%
	\providecommand \bibfnamefont [1]{#1}%
	\providecommand \citenamefont [1]{#1}%
	\providecommand \href@noop [0]{\@secondoftwo}%
	\providecommand \href [0]{\begingroup \@sanitize@url \@href}%
	\providecommand \@href[1]{\@@startlink{#1}\@@href}%
	\providecommand \@@href[1]{\endgroup#1\@@endlink}%
	\providecommand \@sanitize@url [0]{\catcode `\\12\catcode `\$12\catcode
		`\&12\catcode `\#12\catcode `\^12\catcode `\_12\catcode `\%12\relax}%
	\providecommand \@@startlink[1]{}%
	\providecommand \@@endlink[0]{}%
	\providecommand \url  [0]{\begingroup\@sanitize@url \@url }%
	\providecommand \@url [1]{\endgroup\@href {#1}{\urlprefix }}%
	\providecommand \urlprefix  [0]{URL }%
	\providecommand \Eprint [0]{\href }%
	\providecommand \doibase [0]{https://doi.org/}%
	\providecommand \selectlanguage [0]{\@gobble}%
	\providecommand \bibinfo  [0]{\@secondoftwo}%
	\providecommand \bibfield  [0]{\@secondoftwo}%
	\providecommand \translation [1]{[#1]}%
	\providecommand \BibitemOpen [0]{}%
	\providecommand \bibitemStop [0]{}%
	\providecommand \bibitemNoStop [0]{.\EOS\space}%
	\providecommand \EOS [0]{\spacefactor3000\relax}%
	\providecommand \BibitemShut  [1]{\csname bibitem#1\endcsname}%
	\let\auto@bib@innerbib\@empty
	\bibitem [{\citenamefont {Qi}\ and\ \citenamefont {Zhang}(2011)}]{Qi_2011}%
	\BibitemOpen
	\bibfield  {author} {\bibinfo {author} {\bibfnamefont {X.-L.}\ \bibnamefont
			{Qi}}\ and\ \bibinfo {author} {\bibfnamefont {S.-C.}\ \bibnamefont {Zhang}},\
	}\bibfield  {title} {\bibinfo {title} {Topological insulators and
			superconductors},\ }\href {https://doi.org/10.1103/RevModPhys.83.1057}
	{\bibfield  {journal} {\bibinfo  {journal} {Rev. Mod. Phys.}\ }\textbf
		{\bibinfo {volume} {83}},\ \bibinfo {pages} {1057} (\bibinfo {year}
		{2011})}\BibitemShut {NoStop}%
	\bibitem [{\citenamefont {Hasan}\ and\ \citenamefont
		{Kane}(2010)}]{Hasan_2010}%
	\BibitemOpen
	\bibfield  {author} {\bibinfo {author} {\bibfnamefont {M.~Z.}\ \bibnamefont
			{Hasan}}\ and\ \bibinfo {author} {\bibfnamefont {C.~L.}\ \bibnamefont
			{Kane}},\ }\bibfield  {title} {\bibinfo {title} {Colloquium: Topological
			insulators},\ }\href {https://doi.org/10.1103/RevModPhys.82.3045} {\bibfield
		{journal} {\bibinfo  {journal} {Rev. Mod. Phys.}\ }\textbf {\bibinfo {volume}
			{82}},\ \bibinfo {pages} {3045} (\bibinfo {year} {2010})}\BibitemShut
	{NoStop}%
	\bibitem [{\citenamefont {Wan}\ \emph {et~al.}(2011)\citenamefont {Wan},
		\citenamefont {Turner}, \citenamefont {Vishwanath},\ and\ \citenamefont
		{Savrasov}}]{Wan2011}%
	\BibitemOpen
	\bibfield  {author} {\bibinfo {author} {\bibfnamefont {X.}~\bibnamefont
			{Wan}}, \bibinfo {author} {\bibfnamefont {A.~M.}\ \bibnamefont {Turner}},
		\bibinfo {author} {\bibfnamefont {A.}~\bibnamefont {Vishwanath}},\ and\
		\bibinfo {author} {\bibfnamefont {S.~Y.}\ \bibnamefont {Savrasov}},\
	}\bibfield  {title} {\bibinfo {title} {Topological semimetal and fermi-arc
			surface states in the electronic structure of pyrochlore iridates},\ }\href
	{https://doi.org/10.1103/PhysRevB.83.205101} {\bibfield  {journal} {\bibinfo
			{journal} {Phys. Rev. B}\ }\textbf {\bibinfo {volume} {83}},\ \bibinfo
		{pages} {205101} (\bibinfo {year} {2011})}\BibitemShut {NoStop}%
	\bibitem [{\citenamefont {Ahn}\ \emph {et~al.}(2018)\citenamefont {Ahn},
		\citenamefont {Kim}, \citenamefont {Kim},\ and\ \citenamefont
		{Yang}}]{Ahn_2018}%
	\BibitemOpen
	\bibfield  {author} {\bibinfo {author} {\bibfnamefont {J.}~\bibnamefont
			{Ahn}}, \bibinfo {author} {\bibfnamefont {D.}~\bibnamefont {Kim}}, \bibinfo
		{author} {\bibfnamefont {Y.}~\bibnamefont {Kim}},\ and\ \bibinfo {author}
		{\bibfnamefont {B.-J.}\ \bibnamefont {Yang}},\ }\bibfield  {title} {\bibinfo
		{title} {Band topology and linking structure of nodal line semimetals with z2
			monopole charges},\ }\href {https://doi.org/10.1103/PhysRevLett.121.106403}
	{\bibfield  {journal} {\bibinfo  {journal} {Phys. Rev. Lett.}\ }\textbf
		{\bibinfo {volume} {121}},\ \bibinfo {pages} {106403} (\bibinfo {year}
		{2018})}\BibitemShut {NoStop}%
	\bibitem [{\citenamefont {Zhang}\ \emph {et~al.}(2018)\citenamefont {Zhang},
		\citenamefont {Zhu}, \citenamefont {Zhao}, \citenamefont {Yan},\ and\
		\citenamefont {Zhu}}]{Zhang_2018}%
	\BibitemOpen
	\bibfield  {author} {\bibinfo {author} {\bibfnamefont {D.-W.}\ \bibnamefont
			{Zhang}}, \bibinfo {author} {\bibfnamefont {Y.-Q.}\ \bibnamefont {Zhu}},
		\bibinfo {author} {\bibfnamefont {Y.~X.}\ \bibnamefont {Zhao}}, \bibinfo
		{author} {\bibfnamefont {H.}~\bibnamefont {Yan}},\ and\ \bibinfo {author}
		{\bibfnamefont {S.-L.}\ \bibnamefont {Zhu}},\ }\bibfield  {title} {\bibinfo
		{title} {Topological quantum matter with cold atoms},\ }\href
	{https://doi.org/10.1080/00018732.2019.1594094} {\bibfield  {journal}
		{\bibinfo  {journal} {Advances in Physics}\ }\textbf {\bibinfo {volume}
			{67}},\ \bibinfo {pages} {253} (\bibinfo {year} {2018})}\BibitemShut
	{NoStop}%
	\bibitem [{\citenamefont {Yang}\ \emph {et~al.}(2018)\citenamefont {Yang},
		\citenamefont {Yang}, \citenamefont {Derunova}, \citenamefont {Parkin},
		\citenamefont {Yan},\ and\ \citenamefont {Ali}}]{Yang_2018}%
	\BibitemOpen
	\bibfield  {author} {\bibinfo {author} {\bibfnamefont {S.-Y.}\ \bibnamefont
			{Yang}}, \bibinfo {author} {\bibfnamefont {H.}~\bibnamefont {Yang}}, \bibinfo
		{author} {\bibfnamefont {E.}~\bibnamefont {Derunova}}, \bibinfo {author}
		{\bibfnamefont {S.~S.~P.}\ \bibnamefont {Parkin}}, \bibinfo {author}
		{\bibfnamefont {B.}~\bibnamefont {Yan}},\ and\ \bibinfo {author}
		{\bibfnamefont {M.~N.}\ \bibnamefont {Ali}},\ }\bibfield  {title} {\bibinfo
		{title} {Symmetry demanded topological nodal-line materials},\ }\href
	{https://doi.org/10.1080/23746149.2017.1414631} {\bibfield  {journal}
		{\bibinfo  {journal} {Adv. Phys.: X}\ }\textbf {\bibinfo {volume} {3}},\
		\bibinfo {pages} {1414631} (\bibinfo {year} {2018})}\BibitemShut {NoStop}%
	\bibitem [{\citenamefont {Fang}\ \emph {et~al.}(2015)\citenamefont {Fang},
		\citenamefont {Chen}, \citenamefont {Kee},\ and\ \citenamefont
		{Fu}}]{Fang2015}%
	\BibitemOpen
	\bibfield  {author} {\bibinfo {author} {\bibfnamefont {C.}~\bibnamefont
			{Fang}}, \bibinfo {author} {\bibfnamefont {Y.}~\bibnamefont {Chen}}, \bibinfo
		{author} {\bibfnamefont {H.-Y.}\ \bibnamefont {Kee}},\ and\ \bibinfo {author}
		{\bibfnamefont {L.}~\bibnamefont {Fu}},\ }\bibfield  {title} {\bibinfo
		{title} {Topological nodal line semimetals with and without spin-orbital
			coupling},\ }\href {https://doi.org/10.1103/PhysRevB.92.081201} {\bibfield
		{journal} {\bibinfo  {journal} {Phys. Rev. B}\ }\textbf {\bibinfo {volume}
			{92}},\ \bibinfo {pages} {081201(R)} (\bibinfo {year} {2015})}\BibitemShut
	{NoStop}%
	\bibitem [{\citenamefont {Fang}\ \emph {et~al.}(2016)\citenamefont {Fang},
		\citenamefont {Weng}, \citenamefont {Dai},\ and\ \citenamefont
		{Fang}}]{Fang_2016}%
	\BibitemOpen
	\bibfield  {author} {\bibinfo {author} {\bibfnamefont {C.}~\bibnamefont
			{Fang}}, \bibinfo {author} {\bibfnamefont {H.}~\bibnamefont {Weng}}, \bibinfo
		{author} {\bibfnamefont {X.}~\bibnamefont {Dai}},\ and\ \bibinfo {author}
		{\bibfnamefont {Z.}~\bibnamefont {Fang}},\ }\bibfield  {title} {\bibinfo
		{title} {Topological nodal line semimetals},\ }\href
	{https://doi.org/10.1088/1674-1056/25/11/117106} {\bibfield  {journal}
		{\bibinfo  {journal} {Chin. Phys. B}\ }\textbf {\bibinfo {volume} {25}},\
		\bibinfo {pages} {117106} (\bibinfo {year} {2016})}\BibitemShut {NoStop}%
	\bibitem [{\citenamefont {Bian}\ \emph
		{et~al.}(2016{\natexlab{a}})\citenamefont {Bian}, \citenamefont {Chang},
		\citenamefont {Sankar}, \citenamefont {Xu}, \citenamefont {Zheng},
		\citenamefont {Neupert}, \citenamefont {Chiu}, \citenamefont {Huang},
		\citenamefont {Chang}, \citenamefont {Belopolski}, \citenamefont {Sanchez},
		\citenamefont {Neupane}, \citenamefont {Alidoust}, \citenamefont {Liu},
		\citenamefont {Wang}, \citenamefont {Lee}, \citenamefont {Jeng},
		\citenamefont {Zhang}, \citenamefont {Yuan}, \citenamefont {Jia},
		\citenamefont {Bansil}, \citenamefont {Chou}, \citenamefont {Lin},\ and\
		\citenamefont {Hasan}}]{Bian_2016}%
	\BibitemOpen
	\bibfield  {author} {\bibinfo {author} {\bibfnamefont {G.}~\bibnamefont
			{Bian}}, \bibinfo {author} {\bibfnamefont {T.-R.}\ \bibnamefont {Chang}},
		\bibinfo {author} {\bibfnamefont {R.}~\bibnamefont {Sankar}}, \bibinfo
		{author} {\bibfnamefont {S.-Y.}\ \bibnamefont {Xu}}, \bibinfo {author}
		{\bibfnamefont {H.}~\bibnamefont {Zheng}}, \bibinfo {author} {\bibfnamefont
			{T.}~\bibnamefont {Neupert}}, \bibinfo {author} {\bibfnamefont {C.-K.}\
			\bibnamefont {Chiu}}, \bibinfo {author} {\bibfnamefont {S.-M.}\ \bibnamefont
			{Huang}}, \bibinfo {author} {\bibfnamefont {G.}~\bibnamefont {Chang}},
		\bibinfo {author} {\bibfnamefont {I.}~\bibnamefont {Belopolski}}, \bibinfo
		{author} {\bibfnamefont {D.~S.}\ \bibnamefont {Sanchez}}, \bibinfo {author}
		{\bibfnamefont {M.}~\bibnamefont {Neupane}}, \bibinfo {author} {\bibfnamefont
			{N.}~\bibnamefont {Alidoust}}, \bibinfo {author} {\bibfnamefont
			{C.}~\bibnamefont {Liu}}, \bibinfo {author} {\bibfnamefont {B.}~\bibnamefont
			{Wang}}, \bibinfo {author} {\bibfnamefont {C.-C.}\ \bibnamefont {Lee}},
		\bibinfo {author} {\bibfnamefont {H.-T.}\ \bibnamefont {Jeng}}, \bibinfo
		{author} {\bibfnamefont {C.}~\bibnamefont {Zhang}}, \bibinfo {author}
		{\bibfnamefont {Z.}~\bibnamefont {Yuan}}, \bibinfo {author} {\bibfnamefont
			{S.}~\bibnamefont {Jia}}, \bibinfo {author} {\bibfnamefont {A.}~\bibnamefont
			{Bansil}}, \bibinfo {author} {\bibfnamefont {F.}~\bibnamefont {Chou}},
		\bibinfo {author} {\bibfnamefont {H.}~\bibnamefont {Lin}},\ and\ \bibinfo
		{author} {\bibfnamefont {M.~Z.}\ \bibnamefont {Hasan}},\ }\bibfield  {title}
	{\bibinfo {title} {Topological nodal-line fermions in spin-orbit metal
			{PbTaSe}2},\ }\href {https://doi.org/https://doi.org/10.1038/ncomms10556}
	{\bibfield  {journal} {\bibinfo  {journal} {Nat. Commun.}\ }\textbf {\bibinfo
			{volume} {7}},\ \bibinfo {pages} {10556} (\bibinfo {year}
		{2016}{\natexlab{a}})}\BibitemShut {NoStop}%
	\bibitem [{\citenamefont {Bian}\ \emph
		{et~al.}(2016{\natexlab{b}})\citenamefont {Bian}, \citenamefont {Chang},
		\citenamefont {Zheng}, \citenamefont {Velury}, \citenamefont {Xu},
		\citenamefont {Neupert}, \citenamefont {Chiu}, \citenamefont {Huang},
		\citenamefont {Sanchez}, \citenamefont {Belopolski}, \citenamefont
		{Alidoust}, \citenamefont {Chen}, \citenamefont {Chang}, \citenamefont
		{Bansil}, \citenamefont {Jeng}, \citenamefont {Lin},\ and\ \citenamefont
		{Hasan}}]{Bian_20162}%
	\BibitemOpen
	\bibfield  {author} {\bibinfo {author} {\bibfnamefont {G.}~\bibnamefont
			{Bian}}, \bibinfo {author} {\bibfnamefont {T.-R.}\ \bibnamefont {Chang}},
		\bibinfo {author} {\bibfnamefont {H.}~\bibnamefont {Zheng}}, \bibinfo
		{author} {\bibfnamefont {S.}~\bibnamefont {Velury}}, \bibinfo {author}
		{\bibfnamefont {S.-Y.}\ \bibnamefont {Xu}}, \bibinfo {author} {\bibfnamefont
			{T.}~\bibnamefont {Neupert}}, \bibinfo {author} {\bibfnamefont {C.-K.}\
			\bibnamefont {Chiu}}, \bibinfo {author} {\bibfnamefont {S.-M.}\ \bibnamefont
			{Huang}}, \bibinfo {author} {\bibfnamefont {D.~S.}\ \bibnamefont {Sanchez}},
		\bibinfo {author} {\bibfnamefont {I.}~\bibnamefont {Belopolski}}, \bibinfo
		{author} {\bibfnamefont {N.}~\bibnamefont {Alidoust}}, \bibinfo {author}
		{\bibfnamefont {P.-J.}\ \bibnamefont {Chen}}, \bibinfo {author}
		{\bibfnamefont {G.}~\bibnamefont {Chang}}, \bibinfo {author} {\bibfnamefont
			{A.}~\bibnamefont {Bansil}}, \bibinfo {author} {\bibfnamefont {H.-T.}\
			\bibnamefont {Jeng}}, \bibinfo {author} {\bibfnamefont {H.}~\bibnamefont
			{Lin}},\ and\ \bibinfo {author} {\bibfnamefont {M.~Z.}\ \bibnamefont
			{Hasan}},\ }\bibfield  {title} {\bibinfo {title} {Drumhead surface states and
			topological nodal-line fermions {inTlTaSe}2},\ }\href
	{https://doi.org/https://doi.org/10.1103/PhysRevB.93.121113} {\bibfield
		{journal} {\bibinfo  {journal} {Phys. Rev. B}\ }\textbf {\bibinfo {volume}
			{93}},\ \bibinfo {pages} {121113(R)} (\bibinfo {year}
		{2016}{\natexlab{b}})}\BibitemShut {NoStop}%
	\bibitem [{\citenamefont {Du}\ \emph {et~al.}(2017)\citenamefont {Du},
		\citenamefont {Bo}, \citenamefont {Wang}, \citenamefont {Kan}, \citenamefont
		{Duan}, \citenamefont {Savrasov},\ and\ \citenamefont {Wan}}]{Du_2017}%
	\BibitemOpen
	\bibfield  {author} {\bibinfo {author} {\bibfnamefont {Y.}~\bibnamefont
			{Du}}, \bibinfo {author} {\bibfnamefont {X.}~\bibnamefont {Bo}}, \bibinfo
		{author} {\bibfnamefont {D.}~\bibnamefont {Wang}}, \bibinfo {author}
		{\bibfnamefont {E.~J.}\ \bibnamefont {Kan}}, \bibinfo {author} {\bibfnamefont
			{C.-G.}\ \bibnamefont {Duan}}, \bibinfo {author} {\bibfnamefont {S.~Y.}\
			\bibnamefont {Savrasov}},\ and\ \bibinfo {author} {\bibfnamefont
			{X.}~\bibnamefont {Wan}},\ }\bibfield  {title} {\bibinfo {title} {Emergence
			of topological nodal lines and type-ii weyl nodes in the strong spin-orbit
			coupling system $\mathrm{InNb}{X}_{2}$ ($x=\mathrm{S}$,se)},\ }\href
	{https://doi.org/10.1103/PhysRevB.96.235152} {\bibfield  {journal} {\bibinfo
			{journal} {Phys. Rev. B}\ }\textbf {\bibinfo {volume} {96}},\ \bibinfo
		{pages} {235152} (\bibinfo {year} {2017})}\BibitemShut {NoStop}%
	\bibitem [{\citenamefont {Li}\ \emph {et~al.}(2021)\citenamefont {Li},
		\citenamefont {Wu}, \citenamefont {Xu}, \citenamefont {Liu}, \citenamefont
		{Ma}, \citenamefont {Lv}, \citenamefont {Yao}, \citenamefont {Liu},
		\citenamefont {Bai}, \citenamefont {Yang}, \citenamefont {Qiao},
		\citenamefont {Li}, \citenamefont {Li}, \citenamefont {Xing}, \citenamefont
		{Huang}, \citenamefont {Ma}, \citenamefont {Shi}, \citenamefont {Cao},
		\citenamefont {Liu}, \citenamefont {Liu}, \citenamefont {Jia},\ and\
		\citenamefont {Xu}}]{Li_2021}%
	\BibitemOpen
	\bibfield  {author} {\bibinfo {author} {\bibfnamefont {Y.}~\bibnamefont
			{Li}}, \bibinfo {author} {\bibfnamefont {Y.}~\bibnamefont {Wu}}, \bibinfo
		{author} {\bibfnamefont {C.}~\bibnamefont {Xu}}, \bibinfo {author}
		{\bibfnamefont {N.}~\bibnamefont {Liu}}, \bibinfo {author} {\bibfnamefont
			{J.}~\bibnamefont {Ma}}, \bibinfo {author} {\bibfnamefont {B.}~\bibnamefont
			{Lv}}, \bibinfo {author} {\bibfnamefont {G.}~\bibnamefont {Yao}}, \bibinfo
		{author} {\bibfnamefont {Y.}~\bibnamefont {Liu}}, \bibinfo {author}
		{\bibfnamefont {H.}~\bibnamefont {Bai}}, \bibinfo {author} {\bibfnamefont
			{X.}~\bibnamefont {Yang}}, \bibinfo {author} {\bibfnamefont {L.}~\bibnamefont
			{Qiao}}, \bibinfo {author} {\bibfnamefont {M.}~\bibnamefont {Li}}, \bibinfo
		{author} {\bibfnamefont {L.}~\bibnamefont {Li}}, \bibinfo {author}
		{\bibfnamefont {H.}~\bibnamefont {Xing}}, \bibinfo {author} {\bibfnamefont
			{Y.}~\bibnamefont {Huang}}, \bibinfo {author} {\bibfnamefont
			{J.}~\bibnamefont {Ma}}, \bibinfo {author} {\bibfnamefont {M.}~\bibnamefont
			{Shi}}, \bibinfo {author} {\bibfnamefont {C.}~\bibnamefont {Cao}}, \bibinfo
		{author} {\bibfnamefont {Y.}~\bibnamefont {Liu}}, \bibinfo {author}
		{\bibfnamefont {C.}~\bibnamefont {Liu}}, \bibinfo {author} {\bibfnamefont
			{J.}~\bibnamefont {Jia}},\ and\ \bibinfo {author} {\bibfnamefont {Z.-A.}\
			\bibnamefont {Xu}},\ }\bibfield  {title} {\bibinfo {title} {Anisotropic
			gapping of topological weyl rings in the charge-density-wave superconductor
			in {TaSe}$_2$},\ }\href
	{https://doi.org/https://doi.org/10.1016/j.scib.2020.09.007} {\bibfield
		{journal} {\bibinfo  {journal} {Sci. Bull.}\ }\textbf {\bibinfo {volume}
			{66}},\ \bibinfo {pages} {243} (\bibinfo {year} {2021})}\BibitemShut
	{NoStop}%
	\bibitem [{\citenamefont {Li}\ \emph {et~al.}(2020)\citenamefont {Li},
		\citenamefont {Wu}, \citenamefont {Zhou}, \citenamefont {Bu}, \citenamefont
		{Xu}, \citenamefont {Qiao}, \citenamefont {Li}, \citenamefont {Bai},
		\citenamefont {Ma}, \citenamefont {Tao}, \citenamefont {Cao}, \citenamefont
		{Yin},\ and\ \citenamefont {Xu}}]{Li_2020}%
	\BibitemOpen
	\bibfield  {author} {\bibinfo {author} {\bibfnamefont {Y.}~\bibnamefont
			{Li}}, \bibinfo {author} {\bibfnamefont {Z.}~\bibnamefont {Wu}}, \bibinfo
		{author} {\bibfnamefont {J.}~\bibnamefont {Zhou}}, \bibinfo {author}
		{\bibfnamefont {K.}~\bibnamefont {Bu}}, \bibinfo {author} {\bibfnamefont
			{C.}~\bibnamefont {Xu}}, \bibinfo {author} {\bibfnamefont {L.}~\bibnamefont
			{Qiao}}, \bibinfo {author} {\bibfnamefont {M.}~\bibnamefont {Li}}, \bibinfo
		{author} {\bibfnamefont {H.}~\bibnamefont {Bai}}, \bibinfo {author}
		{\bibfnamefont {J.}~\bibnamefont {Ma}}, \bibinfo {author} {\bibfnamefont
			{Q.}~\bibnamefont {Tao}}, \bibinfo {author} {\bibfnamefont {C.}~\bibnamefont
			{Cao}}, \bibinfo {author} {\bibfnamefont {Y.}~\bibnamefont {Yin}},\ and\
		\bibinfo {author} {\bibfnamefont {Z.-A.}\ \bibnamefont {Xu}},\ }\bibfield
	{title} {\bibinfo {title} {Enhanced anisotropic superconductivity in the
			topological nodal-line semimetal {In}$_x${TaS}$_2$},\ }\href
	{https://doi.org/https://doi.org/10.1103/PhysRevB.102.224503} {\bibfield
		{journal} {\bibinfo  {journal} {Phys. Rev. B}\ }\textbf {\bibinfo {volume}
			{102}},\ \bibinfo {pages} {224503} (\bibinfo {year} {2020})}\BibitemShut
	{NoStop}%
	\bibitem [{\citenamefont {Xu}\ \emph {et~al.}(2017)\citenamefont {Xu},
		\citenamefont {Yu}, \citenamefont {Fang}, \citenamefont {Dai},\ and\
		\citenamefont {Weng}}]{Xu_2017}%
	\BibitemOpen
	\bibfield  {author} {\bibinfo {author} {\bibfnamefont {Q.}~\bibnamefont
			{Xu}}, \bibinfo {author} {\bibfnamefont {R.}~\bibnamefont {Yu}}, \bibinfo
		{author} {\bibfnamefont {Z.}~\bibnamefont {Fang}}, \bibinfo {author}
		{\bibfnamefont {X.}~\bibnamefont {Dai}},\ and\ \bibinfo {author}
		{\bibfnamefont {H.}~\bibnamefont {Weng}},\ }\bibfield  {title} {\bibinfo
		{title} {Topological nodal line semimetals in the {CaP}$_3$ family of
			materials},\ }\href {https://doi.org/10.1103/PhysRevB.95.045136} {\bibfield
		{journal} {\bibinfo  {journal} {Phys. Rev. B}\ }\textbf {\bibinfo {volume}
			{95}},\ \bibinfo {pages} {045136} (\bibinfo {year} {2017})}\BibitemShut
	{NoStop}%
	\bibitem [{\citenamefont {Hu}\ \emph {et~al.}(2016)\citenamefont {Hu},
		\citenamefont {Tang}, \citenamefont {Liu}, \citenamefont {Liu}, \citenamefont
		{Zhu}, \citenamefont {Graf}, \citenamefont {Myhro}, \citenamefont {Tran},
		\citenamefont {Lau}, \citenamefont {Wei},\ and\ \citenamefont
		{Mao}}]{Hu_2016}%
	\BibitemOpen
	\bibfield  {author} {\bibinfo {author} {\bibfnamefont {J.}~\bibnamefont
			{Hu}}, \bibinfo {author} {\bibfnamefont {Z.}~\bibnamefont {Tang}}, \bibinfo
		{author} {\bibfnamefont {J.}~\bibnamefont {Liu}}, \bibinfo {author}
		{\bibfnamefont {X.}~\bibnamefont {Liu}}, \bibinfo {author} {\bibfnamefont
			{Y.}~\bibnamefont {Zhu}}, \bibinfo {author} {\bibfnamefont {D.}~\bibnamefont
			{Graf}}, \bibinfo {author} {\bibfnamefont {K.}~\bibnamefont {Myhro}},
		\bibinfo {author} {\bibfnamefont {S.}~\bibnamefont {Tran}}, \bibinfo {author}
		{\bibfnamefont {C.~N.}\ \bibnamefont {Lau}}, \bibinfo {author} {\bibfnamefont
			{J.}~\bibnamefont {Wei}},\ and\ \bibinfo {author} {\bibfnamefont
			{Z.}~\bibnamefont {Mao}},\ }\bibfield  {title} {\bibinfo {title} {Evidence of
			topological nodal-line fermions in {ZrSiSe} and {ZrSiTe}},\ }\href
	{https://doi.org/https://doi.org/10.1103/PhysRevLett.117.016602} {\bibfield
		{journal} {\bibinfo  {journal} {Phys. Rev. Lett.}\ }\textbf {\bibinfo
			{volume} {117}},\ \bibinfo {pages} {016602} (\bibinfo {year}
		{2016})}\BibitemShut {NoStop}%
	\bibitem [{\citenamefont {Chen}\ and\ \citenamefont {Lado}(2019)}]{Chen2019}%
	\BibitemOpen
	\bibfield  {author} {\bibinfo {author} {\bibfnamefont {W.}~\bibnamefont
			{Chen}}\ and\ \bibinfo {author} {\bibfnamefont {J.~L.}\ \bibnamefont
			{Lado}},\ }\bibfield  {title} {\bibinfo {title} {Interaction-driven surface
			chern insulator in nodal line semimetals},\ }\href
	{https://doi.org/10.1103/PhysRevLett.122.016803} {\bibfield  {journal}
		{\bibinfo  {journal} {Phys. Rev. Lett.}\ }\textbf {\bibinfo {volume} {122}},\
		\bibinfo {pages} {016803} (\bibinfo {year} {2019})}\BibitemShut {NoStop}%
	\bibitem [{\citenamefont {Molina}\ and\ \citenamefont
		{Gonz{\'{a}}lez}(2018)}]{Molina_2018}%
	\BibitemOpen
	\bibfield  {author} {\bibinfo {author} {\bibfnamefont {R.~A.}\ \bibnamefont
			{Molina}}\ and\ \bibinfo {author} {\bibfnamefont {J.}~\bibnamefont
			{Gonz{\'{a}}lez}},\ }\bibfield  {title} {\bibinfo {title} {Surface and 3d
			quantum hall effects from engineering of exceptional points in nodal-line
			semimetals},\ }\href {https://doi.org/10.1103/PhysRevLett.120.146601}
	{\bibfield  {journal} {\bibinfo  {journal} {Phys. Rev. Lett.}\ }\textbf
		{\bibinfo {volume} {120}},\ \bibinfo {pages} {146601} (\bibinfo {year}
		{2018})}\BibitemShut {NoStop}%
	\bibitem [{\citenamefont {Wang}\ \emph {et~al.}(2019)\citenamefont {Wang},
		\citenamefont {Wieder}, \citenamefont {Li}, \citenamefont {Yan},\ and\
		\citenamefont {Bernevig}}]{Wang2019}%
	\BibitemOpen
	\bibfield  {author} {\bibinfo {author} {\bibfnamefont {Z.}~\bibnamefont
			{Wang}}, \bibinfo {author} {\bibfnamefont {B.~J.}\ \bibnamefont {Wieder}},
		\bibinfo {author} {\bibfnamefont {J.}~\bibnamefont {Li}}, \bibinfo {author}
		{\bibfnamefont {B.}~\bibnamefont {Yan}},\ and\ \bibinfo {author}
		{\bibfnamefont {B.~A.}\ \bibnamefont {Bernevig}},\ }\bibfield  {title}
	{\bibinfo {title} {Higher-order topology, monopole nodal lines, and the
			origin of large fermi arcs in transition metal dichalcogenides {XTe}$_2$
			($\mathrm{X}=\mathrm{Mo},\mathrm{W}$)},\ }\href
	{https://doi.org/10.1103/PhysRevLett.123.186401} {\bibfield  {journal}
		{\bibinfo  {journal} {Phys. Rev. Lett.}\ }\textbf {\bibinfo {volume} {123}},\
		\bibinfo {pages} {186401} (\bibinfo {year} {2019})}\BibitemShut {NoStop}%
	\bibitem [{\citenamefont {Schindler}\ \emph
		{et~al.}(2018{\natexlab{a}})\citenamefont {Schindler}, \citenamefont {Wang},
		\citenamefont {Vergniory}, \citenamefont {Cook}, \citenamefont {Murani},
		\citenamefont {Sengupta}, \citenamefont {Kasumov}, \citenamefont {Deblock},
		\citenamefont {Jeon}, \citenamefont {Drozdov}, \citenamefont {Bouchiat},
		\citenamefont {Gu{\'{e}}ron}, \citenamefont {Yazdani}, \citenamefont
		{Bernevig},\ and\ \citenamefont {Neupert}}]{Schindler_2018a}%
	\BibitemOpen
	\bibfield  {author} {\bibinfo {author} {\bibfnamefont {F.}~\bibnamefont
			{Schindler}}, \bibinfo {author} {\bibfnamefont {Z.}~\bibnamefont {Wang}},
		\bibinfo {author} {\bibfnamefont {M.~G.}\ \bibnamefont {Vergniory}}, \bibinfo
		{author} {\bibfnamefont {A.~M.}\ \bibnamefont {Cook}}, \bibinfo {author}
		{\bibfnamefont {A.}~\bibnamefont {Murani}}, \bibinfo {author} {\bibfnamefont
			{S.}~\bibnamefont {Sengupta}}, \bibinfo {author} {\bibfnamefont {A.~Y.}\
			\bibnamefont {Kasumov}}, \bibinfo {author} {\bibfnamefont {R.}~\bibnamefont
			{Deblock}}, \bibinfo {author} {\bibfnamefont {S.}~\bibnamefont {Jeon}},
		\bibinfo {author} {\bibfnamefont {I.}~\bibnamefont {Drozdov}}, \bibinfo
		{author} {\bibfnamefont {H.}~\bibnamefont {Bouchiat}}, \bibinfo {author}
		{\bibfnamefont {S.}~\bibnamefont {Gu{\'{e}}ron}}, \bibinfo {author}
		{\bibfnamefont {A.}~\bibnamefont {Yazdani}}, \bibinfo {author} {\bibfnamefont
			{B.~A.}\ \bibnamefont {Bernevig}},\ and\ \bibinfo {author} {\bibfnamefont
			{T.}~\bibnamefont {Neupert}},\ }\bibfield  {title} {\bibinfo {title}
		{Higher-order topology in bismuth},\ }\href
	{https://doi.org/10.1038/s41567-018-0224-7} {\bibfield  {journal} {\bibinfo
			{journal} {Nat. Phys.}\ }\textbf {\bibinfo {volume} {14}},\ \bibinfo {pages}
		{918} (\bibinfo {year} {2018}{\natexlab{a}})}\BibitemShut {NoStop}%
	\bibitem [{\citenamefont {Schindler}\ \emph
		{et~al.}(2018{\natexlab{b}})\citenamefont {Schindler}, \citenamefont {Cook},
		\citenamefont {Vergniory}, \citenamefont {Wang}, \citenamefont {Parkin},
		\citenamefont {Bernevig},\ and\ \citenamefont {Neupert}}]{Schindler_2018b}%
	\BibitemOpen
	\bibfield  {author} {\bibinfo {author} {\bibfnamefont {F.}~\bibnamefont
			{Schindler}}, \bibinfo {author} {\bibfnamefont {A.~M.}\ \bibnamefont {Cook}},
		\bibinfo {author} {\bibfnamefont {M.~G.}\ \bibnamefont {Vergniory}}, \bibinfo
		{author} {\bibfnamefont {Z.}~\bibnamefont {Wang}}, \bibinfo {author}
		{\bibfnamefont {S.~S.~P.}\ \bibnamefont {Parkin}}, \bibinfo {author}
		{\bibfnamefont {B.~A.}\ \bibnamefont {Bernevig}},\ and\ \bibinfo {author}
		{\bibfnamefont {T.}~\bibnamefont {Neupert}},\ }\bibfield  {title} {\bibinfo
		{title} {Higher-order topological insulators},\ }\href
	{https://doi.org/10.1126/sciadv.aat0346} {\bibfield  {journal} {\bibinfo
			{journal} {Sci. Adv.}\ }\textbf {\bibinfo {volume} {4}},\ \bibinfo {pages}
		{eaat0346} (\bibinfo {year} {2018}{\natexlab{b}})}\BibitemShut {NoStop}%
	\bibitem [{\citenamefont {Ahn}\ and\ \citenamefont {Yang}(2020)}]{Ahn_2020}%
	\BibitemOpen
	\bibfield  {author} {\bibinfo {author} {\bibfnamefont {J.}~\bibnamefont
			{Ahn}}\ and\ \bibinfo {author} {\bibfnamefont {B.-J.}\ \bibnamefont {Yang}},\
	}\bibfield  {title} {\bibinfo {title} {Higher-order topological
			superconductivity of spin-polarized fermions},\ }\href
	{https://doi.org/https://doi.org/10.1103/PhysRevResearch.2.012060} {\bibfield
		{journal} {\bibinfo  {journal} {Phys. Rev. Research}\ }\textbf {\bibinfo
			{volume} {2}},\ \bibinfo {pages} {012060(R)} (\bibinfo {year}
		{2020})}\BibitemShut {NoStop}%
	\bibitem [{\citenamefont {Muechler}\ \emph {et~al.}(2019)\citenamefont
		{Muechler}, \citenamefont {Guguchia}, \citenamefont {Orain}, \citenamefont
		{Nuss}, \citenamefont {Schoop}, \citenamefont {Thomale},\ and\ \citenamefont
		{von Rohr}}]{Muechler_2019}%
	\BibitemOpen
	\bibfield  {author} {\bibinfo {author} {\bibfnamefont {L.}~\bibnamefont
			{Muechler}}, \bibinfo {author} {\bibfnamefont {Z.}~\bibnamefont {Guguchia}},
		\bibinfo {author} {\bibfnamefont {J.-C.}\ \bibnamefont {Orain}}, \bibinfo
		{author} {\bibfnamefont {J.}~\bibnamefont {Nuss}}, \bibinfo {author}
		{\bibfnamefont {L.~M.}\ \bibnamefont {Schoop}}, \bibinfo {author}
		{\bibfnamefont {R.}~\bibnamefont {Thomale}},\ and\ \bibinfo {author}
		{\bibfnamefont {F.~O.}\ \bibnamefont {von Rohr}},\ }\bibfield  {title}
	{\bibinfo {title} {Superconducting order parameter of the nodal-line
			semimetal {NaAlSi}},\ }\href
	{https://doi.org/https://doi.org/10.1063/1.5124242} {\bibfield  {journal}
		{\bibinfo  {journal} {{APL} Materials}\ }\textbf {\bibinfo {volume} {7}},\
		\bibinfo {pages} {121103} (\bibinfo {year} {2019})}\BibitemShut {NoStop}%
	\bibitem [{\citenamefont {Cheng}\ \emph {et~al.}(2020)\citenamefont {Cheng},
		\citenamefont {Xia}, \citenamefont {Shi}, \citenamefont {Yu}, \citenamefont
		{Wang}, \citenamefont {Yan}, \citenamefont {Peets}, \citenamefont {Zhu},
		\citenamefont {Su}, \citenamefont {Zhang}, \citenamefont {Dai}, \citenamefont
		{Wang}, \citenamefont {Zou}, \citenamefont {Yu}, \citenamefont {Kou},
		\citenamefont {Yang}, \citenamefont {Zhao}, \citenamefont {Guo},\ and\
		\citenamefont {Li}}]{Cheng_2020}%
	\BibitemOpen
	\bibfield  {author} {\bibinfo {author} {\bibfnamefont {E.}~\bibnamefont
			{Cheng}}, \bibinfo {author} {\bibfnamefont {W.}~\bibnamefont {Xia}}, \bibinfo
		{author} {\bibfnamefont {X.}~\bibnamefont {Shi}}, \bibinfo {author}
		{\bibfnamefont {Z.}~\bibnamefont {Yu}}, \bibinfo {author} {\bibfnamefont
			{L.}~\bibnamefont {Wang}}, \bibinfo {author} {\bibfnamefont {L.}~\bibnamefont
			{Yan}}, \bibinfo {author} {\bibfnamefont {D.~C.}\ \bibnamefont {Peets}},
		\bibinfo {author} {\bibfnamefont {C.}~\bibnamefont {Zhu}}, \bibinfo {author}
		{\bibfnamefont {H.}~\bibnamefont {Su}}, \bibinfo {author} {\bibfnamefont
			{Y.}~\bibnamefont {Zhang}}, \bibinfo {author} {\bibfnamefont
			{D.}~\bibnamefont {Dai}}, \bibinfo {author} {\bibfnamefont {X.}~\bibnamefont
			{Wang}}, \bibinfo {author} {\bibfnamefont {Z.}~\bibnamefont {Zou}}, \bibinfo
		{author} {\bibfnamefont {N.}~\bibnamefont {Yu}}, \bibinfo {author}
		{\bibfnamefont {X.}~\bibnamefont {Kou}}, \bibinfo {author} {\bibfnamefont
			{W.}~\bibnamefont {Yang}}, \bibinfo {author} {\bibfnamefont {W.}~\bibnamefont
			{Zhao}}, \bibinfo {author} {\bibfnamefont {Y.}~\bibnamefont {Guo}},\ and\
		\bibinfo {author} {\bibfnamefont {S.}~\bibnamefont {Li}},\ }\bibfield
	{title} {\bibinfo {title} {Pressure-induced superconductivity and topological
			phase transitions in the topological nodal-line semimetal {SrAs}$_3$},\
	}\href {https://doi.org/10.1038/s41535-020-0240-6} {\bibfield  {journal}
		{\bibinfo  {journal} {npj Quantum Mater.}\ }\textbf {\bibinfo {volume} {5}},\
		\bibinfo {pages} {38} (\bibinfo {year} {2020})}\BibitemShut {NoStop}%
	\bibitem [{\citenamefont {Gao}\ \emph {et~al.}(2020)\citenamefont {Gao},
		\citenamefont {Si}, \citenamefont {Luo}, \citenamefont {Yan}, \citenamefont
		{Jiang}, \citenamefont {Wang}, \citenamefont {Xu}, \citenamefont {Xu},
		\citenamefont {Tong}, \citenamefont {Song}, \citenamefont {Zhu},
		\citenamefont {Lu},\ and\ \citenamefont {Sun}}]{Gao_2020}%
	\BibitemOpen
	\bibfield  {author} {\bibinfo {author} {\bibfnamefont {J.~J.}\ \bibnamefont
			{Gao}}, \bibinfo {author} {\bibfnamefont {J.~G.}\ \bibnamefont {Si}},
		\bibinfo {author} {\bibfnamefont {X.}~\bibnamefont {Luo}}, \bibinfo {author}
		{\bibfnamefont {J.}~\bibnamefont {Yan}}, \bibinfo {author} {\bibfnamefont
			{Z.~Z.}\ \bibnamefont {Jiang}}, \bibinfo {author} {\bibfnamefont
			{W.}~\bibnamefont {Wang}}, \bibinfo {author} {\bibfnamefont {C.~Q.}\
			\bibnamefont {Xu}}, \bibinfo {author} {\bibfnamefont {X.~F.}\ \bibnamefont
			{Xu}}, \bibinfo {author} {\bibfnamefont {P.}~\bibnamefont {Tong}}, \bibinfo
		{author} {\bibfnamefont {W.~H.}\ \bibnamefont {Song}}, \bibinfo {author}
		{\bibfnamefont {X.~B.}\ \bibnamefont {Zhu}}, \bibinfo {author} {\bibfnamefont
			{W.~J.}\ \bibnamefont {Lu}},\ and\ \bibinfo {author} {\bibfnamefont {Y.~P.}\
			\bibnamefont {Sun}},\ }\bibfield  {title} {\bibinfo {title} {Superconducting
			and topological properties in centrosymmetric {PbTaS}$_2$ single crystals},\
	}\href {https://doi.org/10.1021/acs.jpcc.0c00527} {\bibfield  {journal}
		{\bibinfo  {journal} {J. Phys. Chem. C}\ }\textbf {\bibinfo {volume} {124}},\
		\bibinfo {pages} {6349} (\bibinfo {year} {2020})}\BibitemShut {NoStop}%
	\bibitem [{\citenamefont {Aggarwal}\ \emph {et~al.}(2019)\citenamefont
		{Aggarwal}, \citenamefont {Singh}, \citenamefont {Aslam}, \citenamefont
		{Singha}, \citenamefont {Pariari}, \citenamefont {Gayen}, \citenamefont
		{Kabir}, \citenamefont {Mandal},\ and\ \citenamefont
		{Sheet}}]{Aggarwal_2019}%
	\BibitemOpen
	\bibfield  {author} {\bibinfo {author} {\bibfnamefont {L.}~\bibnamefont
			{Aggarwal}}, \bibinfo {author} {\bibfnamefont {C.~K.}\ \bibnamefont {Singh}},
		\bibinfo {author} {\bibfnamefont {M.}~\bibnamefont {Aslam}}, \bibinfo
		{author} {\bibfnamefont {R.}~\bibnamefont {Singha}}, \bibinfo {author}
		{\bibfnamefont {A.}~\bibnamefont {Pariari}}, \bibinfo {author} {\bibfnamefont
			{S.}~\bibnamefont {Gayen}}, \bibinfo {author} {\bibfnamefont
			{M.}~\bibnamefont {Kabir}}, \bibinfo {author} {\bibfnamefont
			{P.}~\bibnamefont {Mandal}},\ and\ \bibinfo {author} {\bibfnamefont
			{G.}~\bibnamefont {Sheet}},\ }\bibfield  {title} {\bibinfo {title}
		{Tip-induced superconductivity coexisting with preserved topological
			properties in line-nodal semimetal {ZrSiS}},\ }\href
	{https://doi.org/10.1088/1361-648X/ab3b61} {\bibfield  {journal} {\bibinfo
			{journal} {J. Phys.: Condens. Matter}\ }\textbf {\bibinfo {volume} {31}},\
		\bibinfo {pages} {485707} (\bibinfo {year} {2019})}\BibitemShut {NoStop}%
	\bibitem [{\citenamefont {Zhang}\ \emph {et~al.}(2016)\citenamefont {Zhang},
		\citenamefont {Yuan}, \citenamefont {Bian}, \citenamefont {Xu}, \citenamefont
		{Zhang}, \citenamefont {Hasan},\ and\ \citenamefont {Jia}}]{Zhang_2016}%
	\BibitemOpen
	\bibfield  {author} {\bibinfo {author} {\bibfnamefont {C.-L.}\ \bibnamefont
			{Zhang}}, \bibinfo {author} {\bibfnamefont {Z.}~\bibnamefont {Yuan}},
		\bibinfo {author} {\bibfnamefont {G.}~\bibnamefont {Bian}}, \bibinfo {author}
		{\bibfnamefont {S.-Y.}\ \bibnamefont {Xu}}, \bibinfo {author} {\bibfnamefont
			{X.}~\bibnamefont {Zhang}}, \bibinfo {author} {\bibfnamefont {M.~Z.}\
			\bibnamefont {Hasan}},\ and\ \bibinfo {author} {\bibfnamefont
			{S.}~\bibnamefont {Jia}},\ }\bibfield  {title} {\bibinfo {title}
		{Superconducting properties in single crystals of the topological nodal
			semimetal {PbTaSe}$_2$},\ }\href {https://doi.org/10.1103/PhysRevB.93.054520}
	{\bibfield  {journal} {\bibinfo  {journal} {Phys. Rev. B}\ }\textbf {\bibinfo
			{volume} {93}},\ \bibinfo {pages} {054520} (\bibinfo {year}
		{2016})}\BibitemShut {NoStop}%
	\bibitem [{\citenamefont {Setty}\ \emph {et~al.}(2017)\citenamefont {Setty},
		\citenamefont {Phillips},\ and\ \citenamefont {Narayan}}]{Setty_2017}%
	\BibitemOpen
	\bibfield  {author} {\bibinfo {author} {\bibfnamefont {C.}~\bibnamefont
			{Setty}}, \bibinfo {author} {\bibfnamefont {P.~W.}\ \bibnamefont
			{Phillips}},\ and\ \bibinfo {author} {\bibfnamefont {A.}~\bibnamefont
			{Narayan}},\ }\bibfield  {title} {\bibinfo {title} {Quasiparticle
			interference and resonant states in normal and superconducting line nodal
			semimetals},\ }\href {https://doi.org/10.1103/PhysRevB.95.140202} {\bibfield
		{journal} {\bibinfo  {journal} {Phys. Rev. B}\ }\textbf {\bibinfo {volume}
			{95}},\ \bibinfo {pages} {140202(R)} (\bibinfo {year} {2017})}\BibitemShut
	{NoStop}%
	\bibitem [{\citenamefont {Nandkishore}(2016)}]{PhysRevB.93.020506}%
	\BibitemOpen
	\bibfield  {author} {\bibinfo {author} {\bibfnamefont {R.}~\bibnamefont
			{Nandkishore}},\ }\bibfield  {title} {\bibinfo {title} {Weyl and dirac loop
			superconductors},\ }\href {https://doi.org/10.1103/PhysRevB.93.020506}
	{\bibfield  {journal} {\bibinfo  {journal} {Phys. Rev. B}\ }\textbf {\bibinfo
			{volume} {93}},\ \bibinfo {pages} {020506(R)} (\bibinfo {year}
		{2016})}\BibitemShut {NoStop}%
	\bibitem [{\citenamefont {Wang}\ and\ \citenamefont
		{Nandkishore}(2017)}]{Wang2017}%
	\BibitemOpen
	\bibfield  {author} {\bibinfo {author} {\bibfnamefont {Y.}~\bibnamefont
			{Wang}}\ and\ \bibinfo {author} {\bibfnamefont {R.~M.}\ \bibnamefont
			{Nandkishore}},\ }\bibfield  {title} {\bibinfo {title} {Topological surface
			superconductivity in doped weyl loop materials},\ }\href
	{https://doi.org/10.1103/PhysRevB.95.060506} {\bibfield  {journal} {\bibinfo
			{journal} {Phys. Rev. B}\ }\textbf {\bibinfo {volume} {95}},\ \bibinfo
		{pages} {060506(R)} (\bibinfo {year} {2017})}\BibitemShut {NoStop}%
	\bibitem [{\citenamefont {Xu}\ \emph {et~al.}(2019)\citenamefont {Xu},
		\citenamefont {Kang}, \citenamefont {Chang}, \citenamefont {Lin},
		\citenamefont {Bian}, \citenamefont {Yuan}, \citenamefont {Qu}, \citenamefont
		{Zhang},\ and\ \citenamefont {Jia}}]{Xu_2019}%
	\BibitemOpen
	\bibfield  {author} {\bibinfo {author} {\bibfnamefont {X.}~\bibnamefont
			{Xu}}, \bibinfo {author} {\bibfnamefont {Z.}~\bibnamefont {Kang}}, \bibinfo
		{author} {\bibfnamefont {T.-R.}\ \bibnamefont {Chang}}, \bibinfo {author}
		{\bibfnamefont {H.}~\bibnamefont {Lin}}, \bibinfo {author} {\bibfnamefont
			{G.}~\bibnamefont {Bian}}, \bibinfo {author} {\bibfnamefont {Z.}~\bibnamefont
			{Yuan}}, \bibinfo {author} {\bibfnamefont {Z.}~\bibnamefont {Qu}}, \bibinfo
		{author} {\bibfnamefont {J.}~\bibnamefont {Zhang}},\ and\ \bibinfo {author}
		{\bibfnamefont {S.}~\bibnamefont {Jia}},\ }\bibfield  {title} {\bibinfo
		{title} {Quantum oscillations in the noncentrosymmetric superconductor and
			topological nodal-line semimetal {PbTaSe}$_2$},\ }\href
	{https://doi.org/10.1103/PhysRevB.99.104516} {\bibfield  {journal} {\bibinfo
			{journal} {Phys. Rev. B}\ }\textbf {\bibinfo {volume} {99}},\ \bibinfo
		{pages} {104516} (\bibinfo {year} {2019})}\BibitemShut {NoStop}%
	\bibitem [{\citenamefont {Chen}\ \emph {et~al.}(2019)\citenamefont {Chen},
		\citenamefont {Wu}, \citenamefont {Jin}, \citenamefont {Li}, \citenamefont
		{Wang}, \citenamefont {Duan}, \citenamefont {Han}, \citenamefont {Li},
		\citenamefont {Long}, \citenamefont {Zhang}, \citenamefont {Chen},\ and\
		\citenamefont {Teng}}]{Chen_2019}%
	\BibitemOpen
	\bibfield  {author} {\bibinfo {author} {\bibfnamefont {D.-Y.}\ \bibnamefont
			{Chen}}, \bibinfo {author} {\bibfnamefont {Y.}~\bibnamefont {Wu}}, \bibinfo
		{author} {\bibfnamefont {L.}~\bibnamefont {Jin}}, \bibinfo {author}
		{\bibfnamefont {Y.}~\bibnamefont {Li}}, \bibinfo {author} {\bibfnamefont
			{X.}~\bibnamefont {Wang}}, \bibinfo {author} {\bibfnamefont {J.}~\bibnamefont
			{Duan}}, \bibinfo {author} {\bibfnamefont {J.}~\bibnamefont {Han}}, \bibinfo
		{author} {\bibfnamefont {X.}~\bibnamefont {Li}}, \bibinfo {author}
		{\bibfnamefont {Y.-Z.}\ \bibnamefont {Long}}, \bibinfo {author}
		{\bibfnamefont {X.}~\bibnamefont {Zhang}}, \bibinfo {author} {\bibfnamefont
			{D.}~\bibnamefont {Chen}},\ and\ \bibinfo {author} {\bibfnamefont
			{B.}~\bibnamefont {Teng}},\ }\bibfield  {title} {\bibinfo {title}
		{Superconducting properties in a candidate topological nodal line semimetal
			{SnTaS}$_2$ with a centrosymmetric crystal structure},\ }\href
	{https://doi.org/10.1103/PhysRevB.100.064516} {\bibfield  {journal} {\bibinfo
			{journal} {Phys. Rev. B}\ }\textbf {\bibinfo {volume} {100}},\ \bibinfo
		{pages} {064516} (\bibinfo {year} {2019})}\BibitemShut {NoStop}%
	\bibitem [{\citenamefont {Schnyder}\ and\ \citenamefont
		{Brydon}(2015)}]{Schnyder_2015}%
	\BibitemOpen
	\bibfield  {author} {\bibinfo {author} {\bibfnamefont {A.~P.}\ \bibnamefont
			{Schnyder}}\ and\ \bibinfo {author} {\bibfnamefont {P.~M.~R.}\ \bibnamefont
			{Brydon}},\ }\bibfield  {title} {\bibinfo {title} {Topological surface states
			in nodal superconductors},\ }\href
	{https://doi.org/10.1088/0953-8984/27/24/243201} {\bibfield  {journal}
		{\bibinfo  {journal} {J. Phys.: Condens. Matter}\ }\textbf {\bibinfo {volume}
			{27}},\ \bibinfo {pages} {243201} (\bibinfo {year} {2015})}\BibitemShut
	{NoStop}%
	\bibitem [{\citenamefont {Shapourian}\ \emph {et~al.}(2018)\citenamefont
		{Shapourian}, \citenamefont {Wang},\ and\ \citenamefont
		{Ryu}}]{Shapourian2018}%
	\BibitemOpen
	\bibfield  {author} {\bibinfo {author} {\bibfnamefont {H.}~\bibnamefont
			{Shapourian}}, \bibinfo {author} {\bibfnamefont {Y.}~\bibnamefont {Wang}},\
		and\ \bibinfo {author} {\bibfnamefont {S.}~\bibnamefont {Ryu}},\ }\bibfield
	{title} {\bibinfo {title} {Topological crystalline superconductivity and
			second-order topological superconductivity in nodal-loop materials},\ }\href
	{https://doi.org/10.1103/PhysRevB.97.094508} {\bibfield  {journal} {\bibinfo
			{journal} {Phys. Rev. B}\ }\textbf {\bibinfo {volume} {97}},\ \bibinfo
		{pages} {094508} (\bibinfo {year} {2018})}\BibitemShut {NoStop}%
	\bibitem [{\citenamefont {Po}\ \emph {et~al.}(2018)\citenamefont {Po},
		\citenamefont {Watanabe},\ and\ \citenamefont {Vishwanath}}]{Po_2018}%
	\BibitemOpen
	\bibfield  {author} {\bibinfo {author} {\bibfnamefont {H.~C.}\ \bibnamefont
			{Po}}, \bibinfo {author} {\bibfnamefont {H.}~\bibnamefont {Watanabe}},\ and\
		\bibinfo {author} {\bibfnamefont {A.}~\bibnamefont {Vishwanath}},\ }\bibfield
	{title} {\bibinfo {title} {Fragile topology and wannier obstructions},\
	}\href {https://doi.org/10.1103/PhysRevLett.121.126402} {\bibfield  {journal}
		{\bibinfo  {journal} {Phys. Rev. Lett.}\ }\textbf {\bibinfo {volume} {121}},\
		\bibinfo {pages} {126402} (\bibinfo {year} {2018})}\BibitemShut {NoStop}%
	\bibitem [{\citenamefont {Ahn}\ \emph {et~al.}(2019)\citenamefont {Ahn},
		\citenamefont {Park},\ and\ \citenamefont {Yang}}]{Ahn_2019}%
	\BibitemOpen
	\bibfield  {author} {\bibinfo {author} {\bibfnamefont {J.}~\bibnamefont
			{Ahn}}, \bibinfo {author} {\bibfnamefont {S.}~\bibnamefont {Park}},\ and\
		\bibinfo {author} {\bibfnamefont {B.-J.}\ \bibnamefont {Yang}},\ }\bibfield
	{title} {\bibinfo {title} {Failure of nielsen-ninomiya theorem and fragile
			topology in two-dimensional systems with space-time inversion symmetry:
			Application to twisted bilayer graphene at magic angle},\ }\href
	{https://doi.org/10.1103/PhysRevX.9.021013} {\bibfield  {journal} {\bibinfo
			{journal} {Phys. Rev. X}\ }\textbf {\bibinfo {volume} {9}},\ \bibinfo {pages}
		{021013} (\bibinfo {year} {2019})}\BibitemShut {NoStop}%
	\bibitem [{\citenamefont {Bradlyn}\ \emph {et~al.}(2019)\citenamefont
		{Bradlyn}, \citenamefont {Wang}, \citenamefont {Cano},\ and\ \citenamefont
		{Bernevig}}]{Bradlyn_2019}%
	\BibitemOpen
	\bibfield  {author} {\bibinfo {author} {\bibfnamefont {B.}~\bibnamefont
			{Bradlyn}}, \bibinfo {author} {\bibfnamefont {Z.}~\bibnamefont {Wang}},
		\bibinfo {author} {\bibfnamefont {J.}~\bibnamefont {Cano}},\ and\ \bibinfo
		{author} {\bibfnamefont {B.~A.}\ \bibnamefont {Bernevig}},\ }\bibfield
	{title} {\bibinfo {title} {Disconnected elementary band representations,
			fragile topology, and wilson loops as topological indices: An example on the
			triangular lattice},\ }\href {https://doi.org/10.1103/PhysRevB.99.045140}
	{\bibfield  {journal} {\bibinfo  {journal} {Phys. Rev. B}\ }\textbf {\bibinfo
			{volume} {99}},\ \bibinfo {pages} {045140} (\bibinfo {year}
		{2019})}\BibitemShut {NoStop}%
	\bibitem [{\citenamefont {Hwang}\ \emph {et~al.}(2019)\citenamefont {Hwang},
		\citenamefont {Ahn},\ and\ \citenamefont {Yang}}]{Hwang_2019}%
	\BibitemOpen
	\bibfield  {author} {\bibinfo {author} {\bibfnamefont {Y.}~\bibnamefont
			{Hwang}}, \bibinfo {author} {\bibfnamefont {J.}~\bibnamefont {Ahn}},\ and\
		\bibinfo {author} {\bibfnamefont {B.-J.}\ \bibnamefont {Yang}},\ }\bibfield
	{title} {\bibinfo {title} {Fragile topology protected by inversion symmetry:
			Diagnosis, bulk-boundary correspondence, and wilson loop},\ }\href
	{https://doi.org/10.1103/PhysRevB.100.205126} {\bibfield  {journal} {\bibinfo
			{journal} {Phys. Rev. B}\ }\textbf {\bibinfo {volume} {100}},\ \bibinfo
		{pages} {205126} (\bibinfo {year} {2019})}\BibitemShut {NoStop}%
	\bibitem [{\citenamefont {Alexandradinata}\ \emph {et~al.}(2020)\citenamefont
		{Alexandradinata}, \citenamefont {H\"oller}, \citenamefont {Wang},
		\citenamefont {Cheng},\ and\ \citenamefont {Lu}}]{Alexandradinata_2020}%
	\BibitemOpen
	\bibfield  {author} {\bibinfo {author} {\bibfnamefont {A.}~\bibnamefont
			{Alexandradinata}}, \bibinfo {author} {\bibfnamefont {J.}~\bibnamefont
			{H\"oller}}, \bibinfo {author} {\bibfnamefont {C.}~\bibnamefont {Wang}},
		\bibinfo {author} {\bibfnamefont {H.}~\bibnamefont {Cheng}},\ and\ \bibinfo
		{author} {\bibfnamefont {L.}~\bibnamefont {Lu}},\ }\bibfield  {title}
	{\bibinfo {title} {Crystallographic splitting theorem for band
			representations and fragile topological photonic crystals},\ }\href
	{https://doi.org/10.1103/PhysRevB.102.115117} {\bibfield  {journal} {\bibinfo
			{journal} {Phys. Rev. B}\ }\textbf {\bibinfo {volume} {102}},\ \bibinfo
		{pages} {115117} (\bibinfo {year} {2020})}\BibitemShut {NoStop}%
	\bibitem [{\citenamefont {Cano}\ \emph {et~al.}(2018)\citenamefont {Cano},
		\citenamefont {Bradlyn}, \citenamefont {Wang}, \citenamefont {Elcoro},
		\citenamefont {Vergniory}, \citenamefont {Felser}, \citenamefont {Aroyo},\
		and\ \citenamefont {Bernevig}}]{Cano_2018}%
	\BibitemOpen
	\bibfield  {author} {\bibinfo {author} {\bibfnamefont {J.}~\bibnamefont
			{Cano}}, \bibinfo {author} {\bibfnamefont {B.}~\bibnamefont {Bradlyn}},
		\bibinfo {author} {\bibfnamefont {Z.}~\bibnamefont {Wang}}, \bibinfo {author}
		{\bibfnamefont {L.}~\bibnamefont {Elcoro}}, \bibinfo {author} {\bibfnamefont
			{M.~G.}\ \bibnamefont {Vergniory}}, \bibinfo {author} {\bibfnamefont
			{C.}~\bibnamefont {Felser}}, \bibinfo {author} {\bibfnamefont {M.~I.}\
			\bibnamefont {Aroyo}},\ and\ \bibinfo {author} {\bibfnamefont {B.~A.}\
			\bibnamefont {Bernevig}},\ }\bibfield  {title} {\bibinfo {title} {Topology of
			disconnected elementary band representations},\ }\href
	{https://doi.org/10.1103/PhysRevLett.120.266401} {\bibfield  {journal}
		{\bibinfo  {journal} {Phys. Rev. Lett.}\ }\textbf {\bibinfo {volume} {120}},\
		\bibinfo {pages} {266401} (\bibinfo {year} {2018})}\BibitemShut {NoStop}%
	\bibitem [{\citenamefont {Song}\ \emph
		{et~al.}(2020{\natexlab{a}})\citenamefont {Song}, \citenamefont {Elcoro},\
		and\ \citenamefont {Bernevig}}]{Song_2020a}%
	\BibitemOpen
	\bibfield  {author} {\bibinfo {author} {\bibfnamefont {Z.-D.}\ \bibnamefont
			{Song}}, \bibinfo {author} {\bibfnamefont {L.}~\bibnamefont {Elcoro}},\ and\
		\bibinfo {author} {\bibfnamefont {B.~A.}\ \bibnamefont {Bernevig}},\
	}\bibfield  {title} {\bibinfo {title} {Twisted bulk-boundary correspondence
			of fragile topology},\ }\href {https://doi.org/10.1126/science.aaz7650}
	{\bibfield  {journal} {\bibinfo  {journal} {Science}\ }\textbf {\bibinfo
			{volume} {367}},\ \bibinfo {pages} {794} (\bibinfo {year}
		{2020}{\natexlab{a}})}\BibitemShut {NoStop}%
	\bibitem [{\citenamefont {Song}\ \emph
		{et~al.}(2020{\natexlab{b}})\citenamefont {Song}, \citenamefont {Elcoro},
		\citenamefont {Xu}, \citenamefont {Regnault},\ and\ \citenamefont
		{Bernevig}}]{Song_2020b}%
	\BibitemOpen
	\bibfield  {author} {\bibinfo {author} {\bibfnamefont {Z.-D.}\ \bibnamefont
			{Song}}, \bibinfo {author} {\bibfnamefont {L.}~\bibnamefont {Elcoro}},
		\bibinfo {author} {\bibfnamefont {Y.-F.}\ \bibnamefont {Xu}}, \bibinfo
		{author} {\bibfnamefont {N.}~\bibnamefont {Regnault}},\ and\ \bibinfo
		{author} {\bibfnamefont {B.~A.}\ \bibnamefont {Bernevig}},\ }\bibfield
	{title} {\bibinfo {title} {Fragile phases as affine monoids: Classification
			and material examples},\ }\href {https://doi.org/10.1103/PhysRevX.10.031001}
	{\bibfield  {journal} {\bibinfo  {journal} {Phys. Rev. X}\ }\textbf {\bibinfo
			{volume} {10}},\ \bibinfo {pages} {031001} (\bibinfo {year}
		{2020}{\natexlab{b}})}\BibitemShut {NoStop}%
	\bibitem [{\citenamefont {Wieder}\ and\ \citenamefont
		{Bernevig}(2018)}]{Wieder2018}%
	\BibitemOpen
	\bibfield  {author} {\bibinfo {author} {\bibfnamefont {B.~J.}\ \bibnamefont
			{Wieder}}\ and\ \bibinfo {author} {\bibfnamefont {B.~A.}\ \bibnamefont
			{Bernevig}},\ }\href@noop {} {\bibinfo {title} {The axion insulator as a pump
			of fragile topology}} (\bibinfo {year} {2018}),\ \Eprint
	{https://arxiv.org/abs/1810.02373} {arXiv:1810.02373} \BibitemShut {NoStop}%
	\bibitem [{\citenamefont {Bouhon}\ \emph {et~al.}(2019)\citenamefont {Bouhon},
		\citenamefont {Black-Schaffer},\ and\ \citenamefont {Slager}}]{Bouhon_2019}%
	\BibitemOpen
	\bibfield  {author} {\bibinfo {author} {\bibfnamefont {A.}~\bibnamefont
			{Bouhon}}, \bibinfo {author} {\bibfnamefont {A.~M.}\ \bibnamefont
			{Black-Schaffer}},\ and\ \bibinfo {author} {\bibfnamefont {R.-J.}\
			\bibnamefont {Slager}},\ }\bibfield  {title} {\bibinfo {title} {Wilson loop
			approach to fragile topology of split elementary band representations and
			topological crystalline insulators with time-reversal symmetry},\ }\href
	{https://doi.org/10.1103/PhysRevB.100.195135} {\bibfield  {journal} {\bibinfo
			{journal} {Phys. Rev. B}\ }\textbf {\bibinfo {volume} {100}},\ \bibinfo
		{pages} {195135} (\bibinfo {year} {2019})}\BibitemShut {NoStop}%
	\bibitem [{\citenamefont {Lian}\ \emph {et~al.}(2020)\citenamefont {Lian},
		\citenamefont {Xie},\ and\ \citenamefont {Bernevig}}]{Lian_2020}%
	\BibitemOpen
	\bibfield  {author} {\bibinfo {author} {\bibfnamefont {B.}~\bibnamefont
			{Lian}}, \bibinfo {author} {\bibfnamefont {F.}~\bibnamefont {Xie}},\ and\
		\bibinfo {author} {\bibfnamefont {B.~A.}\ \bibnamefont {Bernevig}},\
	}\bibfield  {title} {\bibinfo {title} {Landau level of fragile topology},\
	}\href {https://doi.org/10.1103/PhysRevB.102.041402} {\bibfield  {journal}
		{\bibinfo  {journal} {Phys. Rev. B}\ }\textbf {\bibinfo {volume} {102}},\
		\bibinfo {pages} {041402(R)} (\bibinfo {year} {2020})}\BibitemShut {NoStop}%
	\bibitem [{\citenamefont {Wieder}\ \emph {et~al.}(2020)\citenamefont {Wieder},
		\citenamefont {Wang}, \citenamefont {Cano}, \citenamefont {Dai},
		\citenamefont {Schoop}, \citenamefont {Bradlyn},\ and\ \citenamefont
		{Bernevig}}]{Wieder_2020}%
	\BibitemOpen
	\bibfield  {author} {\bibinfo {author} {\bibfnamefont {B.~J.}\ \bibnamefont
			{Wieder}}, \bibinfo {author} {\bibfnamefont {Z.}~\bibnamefont {Wang}},
		\bibinfo {author} {\bibfnamefont {J.}~\bibnamefont {Cano}}, \bibinfo {author}
		{\bibfnamefont {X.}~\bibnamefont {Dai}}, \bibinfo {author} {\bibfnamefont
			{L.~M.}\ \bibnamefont {Schoop}}, \bibinfo {author} {\bibfnamefont
			{B.}~\bibnamefont {Bradlyn}},\ and\ \bibinfo {author} {\bibfnamefont {B.~A.}\
			\bibnamefont {Bernevig}},\ }\bibfield  {title} {\bibinfo {title} {Strong and
			fragile topological dirac semimetals with higher-order fermi arcs},\ }\href
	{https://doi.org/10.1038/s41467-020-14443-5} {\bibfield  {journal} {\bibinfo
			{journal} {Nat. Commun.}\ }\textbf {\bibinfo {volume} {11}},\ \bibinfo
		{pages} {627} (\bibinfo {year} {2020})}\BibitemShut {NoStop}%
	\bibitem [{\citenamefont {Slager}\ \emph {et~al.}(2013)\citenamefont {Slager},
		\citenamefont {Mesaros}, \citenamefont {Juricic},\ and\ \citenamefont
		{Zaanen}}]{Slager2013}%
	\BibitemOpen
	\bibfield  {author} {\bibinfo {author} {\bibfnamefont {R.-J.}\ \bibnamefont
			{Slager}}, \bibinfo {author} {\bibfnamefont {A.}~\bibnamefont {Mesaros}},
		\bibinfo {author} {\bibfnamefont {V.}~\bibnamefont {Juri$\mathrm{\check{c}}$i$\mathrm{\acute{c}}$}},\ and\ \bibinfo
		{author} {\bibfnamefont {J.}~\bibnamefont {Zaanen}},\ }\bibfield  {title}
	{\bibinfo {title} {The space group classification of topological
			band-insulators},\ }\href {https://doi.org/10.1038/nphys2513} {\bibfield
		{journal} {\bibinfo  {journal} {Nat. Phys.}\ }\textbf {\bibinfo {volume}
			{9}},\ \bibinfo {pages} {98} (\bibinfo {year} {2013})}\BibitemShut {NoStop}%
	\bibitem [{\citenamefont {Kruthoff}\ \emph {et~al.}(2017)\citenamefont
		{Kruthoff}, \citenamefont {de~Boer}, \citenamefont {van Wezel}, \citenamefont
		{Kane},\ and\ \citenamefont {Slager}}]{PhysRevX.7.041069}%
	\BibitemOpen
	\bibfield  {author} {\bibinfo {author} {\bibfnamefont {J.}~\bibnamefont
			{Kruthoff}}, \bibinfo {author} {\bibfnamefont {J.}~\bibnamefont {de~Boer}},
		\bibinfo {author} {\bibfnamefont {J.}~\bibnamefont {van Wezel}}, \bibinfo
		{author} {\bibfnamefont {C.~L.}\ \bibnamefont {Kane}},\ and\ \bibinfo
		{author} {\bibfnamefont {R.-J.}\ \bibnamefont {Slager}},\ }\bibfield  {title}
	{\bibinfo {title} {Topological classification of crystalline insulators
			through band structure combinatorics},\ }\href
	{https://doi.org/10.1103/PhysRevX.7.041069} {\bibfield  {journal} {\bibinfo
			{journal} {Phys. Rev. X}\ }\textbf {\bibinfo {volume} {7}},\ \bibinfo {pages}
		{041069} (\bibinfo {year} {2017})}\BibitemShut {NoStop}%
	\bibitem [{\citenamefont {Zhang}\ \emph {et~al.}(2019)\citenamefont {Zhang},
		\citenamefont {Jiang}, \citenamefont {Song}, \citenamefont {Huang},
		\citenamefont {He}, \citenamefont {Fang}, \citenamefont {Weng},\ and\
		\citenamefont {Fang}}]{Zhang_2019}%
	\BibitemOpen
	\bibfield  {author} {\bibinfo {author} {\bibfnamefont {T.}~\bibnamefont
			{Zhang}}, \bibinfo {author} {\bibfnamefont {Y.}~\bibnamefont {Jiang}},
		\bibinfo {author} {\bibfnamefont {Z.}~\bibnamefont {Song}}, \bibinfo {author}
		{\bibfnamefont {H.}~\bibnamefont {Huang}}, \bibinfo {author} {\bibfnamefont
			{Y.}~\bibnamefont {He}}, \bibinfo {author} {\bibfnamefont {Z.}~\bibnamefont
			{Fang}}, \bibinfo {author} {\bibfnamefont {H.}~\bibnamefont {Weng}},\ and\
		\bibinfo {author} {\bibfnamefont {C.}~\bibnamefont {Fang}},\ }\bibfield
	{title} {\bibinfo {title} {Catalogue of topological electronic materials},\
	}\href {https://doi.org/10.1038/s41586-019-0944-6} {\bibfield  {journal}
		{\bibinfo  {journal} {Nature}\ }\textbf {\bibinfo {volume} {566}},\ \bibinfo
		{pages} {475} (\bibinfo {year} {2019})}\BibitemShut {NoStop}%
	\bibitem [{\citenamefont {Vergniory}\ \emph {et~al.}(2019)\citenamefont
		{Vergniory}, \citenamefont {Elcoro}, \citenamefont {Felser}, \citenamefont
		{Regnault}, \citenamefont {Bernevig},\ and\ \citenamefont
		{Wang}}]{Vergniory_2019}%
	\BibitemOpen
	\bibfield  {author} {\bibinfo {author} {\bibfnamefont {M.~G.}\ \bibnamefont
			{Vergniory}}, \bibinfo {author} {\bibfnamefont {L.}~\bibnamefont {Elcoro}},
		\bibinfo {author} {\bibfnamefont {C.}~\bibnamefont {Felser}}, \bibinfo
		{author} {\bibfnamefont {N.}~\bibnamefont {Regnault}}, \bibinfo {author}
		{\bibfnamefont {B.~A.}\ \bibnamefont {Bernevig}},\ and\ \bibinfo {author}
		{\bibfnamefont {Z.}~\bibnamefont {Wang}},\ }\bibfield  {title} {\bibinfo
		{title} {A complete catalogue of high-quality topological materials},\ }\href
	{https://doi.org/10.1038/s41586-019-0954-4} {\bibfield  {journal} {\bibinfo
			{journal} {Nature}\ }\textbf {\bibinfo {volume} {566}},\ \bibinfo {pages}
		{480} (\bibinfo {year} {2019})}\BibitemShut {NoStop}%
	\bibitem [{\citenamefont {Bradlyn}\ \emph {et~al.}(2017)\citenamefont
		{Bradlyn}, \citenamefont {Elcoro}, \citenamefont {Cano}, \citenamefont
		{Vergniory}, \citenamefont {Wang}, \citenamefont {Felser}, \citenamefont
		{Aroyo},\ and\ \citenamefont {Bernevig}}]{Bradlyn_2017}%
	\BibitemOpen
	\bibfield  {author} {\bibinfo {author} {\bibfnamefont {B.}~\bibnamefont
			{Bradlyn}}, \bibinfo {author} {\bibfnamefont {L.}~\bibnamefont {Elcoro}},
		\bibinfo {author} {\bibfnamefont {J.}~\bibnamefont {Cano}}, \bibinfo {author}
		{\bibfnamefont {M.~G.}\ \bibnamefont {Vergniory}}, \bibinfo {author}
		{\bibfnamefont {Z.}~\bibnamefont {Wang}}, \bibinfo {author} {\bibfnamefont
			{C.}~\bibnamefont {Felser}}, \bibinfo {author} {\bibfnamefont {M.~I.}\
			\bibnamefont {Aroyo}},\ and\ \bibinfo {author} {\bibfnamefont {B.~A.}\
			\bibnamefont {Bernevig}},\ }\bibfield  {title} {\bibinfo {title} {Topological
			quantum chemistry},\ }\href {https://doi.org/10.1038/nature23268} {\bibfield
		{journal} {\bibinfo  {journal} {Nature}\ }\textbf {\bibinfo {volume} {547}},\
		\bibinfo {pages} {298} (\bibinfo {year} {2017})}\BibitemShut {NoStop}%
	\bibitem [{\citenamefont {Po}\ \emph {et~al.}(2017)\citenamefont {Po},
		\citenamefont {Vishwanath},\ and\ \citenamefont {Watanabe}}]{Po_2017}%
	\BibitemOpen
	\bibfield  {author} {\bibinfo {author} {\bibfnamefont {H.~C.}\ \bibnamefont
			{Po}}, \bibinfo {author} {\bibfnamefont {A.}~\bibnamefont {Vishwanath}},\
		and\ \bibinfo {author} {\bibfnamefont {H.}~\bibnamefont {Watanabe}},\
	}\bibfield  {title} {\bibinfo {title} {Symmetry-based indicators of band
			topology in the 230 space groups},\ }\href
	{https://doi.org/10.1038/s41467-017-00133-2} {\bibfield  {journal} {\bibinfo
			{journal} {Nat. Commun.}\ }\textbf {\bibinfo {volume} {8}},\ \bibinfo {pages}
		{50} (\bibinfo {year} {2017})}\BibitemShut {NoStop}%
	\bibitem [{\citenamefont {Tang}\ \emph {et~al.}(2019)\citenamefont {Tang},
		\citenamefont {Po}, \citenamefont {Vishwanath},\ and\ \citenamefont
		{Wan}}]{Tang_2019}%
	\BibitemOpen
	\bibfield  {author} {\bibinfo {author} {\bibfnamefont {F.}~\bibnamefont
			{Tang}}, \bibinfo {author} {\bibfnamefont {H.~C.}\ \bibnamefont {Po}},
		\bibinfo {author} {\bibfnamefont {A.}~\bibnamefont {Vishwanath}},\ and\
		\bibinfo {author} {\bibfnamefont {X.}~\bibnamefont {Wan}},\ }\bibfield
	{title} {\bibinfo {title} {Comprehensive search for topological materials
			using symmetry indicators},\ }\href
	{https://doi.org/10.1038/s41586-019-0937-5} {\bibfield  {journal} {\bibinfo
			{journal} {Nature}\ }\textbf {\bibinfo {volume} {566}},\ \bibinfo {pages}
		{486} (\bibinfo {year} {2019})}\BibitemShut {NoStop}%
	\bibitem [{\citenamefont {Huang}\ and\ \citenamefont {Hsu}(2021)}]{Huang_2021}%
	\BibitemOpen
	\bibfield  {author} {\bibinfo {author} {\bibfnamefont {S.-J.}\ \bibnamefont
			{Huang}}\ and\ \bibinfo {author} {\bibfnamefont {Y.-T.}\ \bibnamefont
			{Hsu}},\ }\bibfield  {title} {\bibinfo {title} {Faithful derivation of
			symmetry indicators: A case study for topological superconductors with
			time-reversal and inversion symmetries},\ }\href
	{https://doi.org/10.1103/PhysRevResearch.3.013243} {\bibfield  {journal}
		{\bibinfo  {journal} {Phys. Rev. Research}\ }\textbf {\bibinfo {volume}
			{3}},\ \bibinfo {pages} {013243} (\bibinfo {year} {2021})}\BibitemShut
	{NoStop}%
	\bibitem [{\citenamefont {Skurativska}\ \emph {et~al.}(2020)\citenamefont
		{Skurativska}, \citenamefont {Neupert},\ and\ \citenamefont
		{Fischer}}]{Skurativska_2020}%
	\BibitemOpen
	\bibfield  {author} {\bibinfo {author} {\bibfnamefont {A.}~\bibnamefont
			{Skurativska}}, \bibinfo {author} {\bibfnamefont {T.}~\bibnamefont
			{Neupert}},\ and\ \bibinfo {author} {\bibfnamefont {M.~H.}\ \bibnamefont
			{Fischer}},\ }\bibfield  {title} {\bibinfo {title} {Atomic limit and
			inversion-symmetry indicators for topological superconductors},\ }\href
	{https://doi.org/10.1103/PhysRevResearch.2.013064} {\bibfield  {journal}
		{\bibinfo  {journal} {Phys. Rev. Research}\ }\textbf {\bibinfo {volume}
			{2}},\ \bibinfo {pages} {013064} (\bibinfo {year} {2020})}\BibitemShut
	{NoStop}%
	\bibitem [{\citenamefont {Ono}\ \emph {et~al.}(2019)\citenamefont {Ono},
		\citenamefont {Yanase},\ and\ \citenamefont {Watanabe}}]{Ono_2019}%
	\BibitemOpen
	\bibfield  {author} {\bibinfo {author} {\bibfnamefont {S.}~\bibnamefont
			{Ono}}, \bibinfo {author} {\bibfnamefont {Y.}~\bibnamefont {Yanase}},\ and\
		\bibinfo {author} {\bibfnamefont {H.}~\bibnamefont {Watanabe}},\ }\bibfield
	{title} {\bibinfo {title} {Symmetry indicators for topological
			superconductors},\ }\href {https://doi.org/10.1103/PhysRevResearch.1.013012}
	{\bibfield  {journal} {\bibinfo  {journal} {Phys. Rev. Research}\ }\textbf
		{\bibinfo {volume} {1}},\ \bibinfo {pages} {013012} (\bibinfo {year}
		{2019})}\BibitemShut {NoStop}%
	\bibitem [{\citenamefont {Sumita}\ \emph {et~al.}(2019)\citenamefont {Sumita},
		\citenamefont {Nomoto}, \citenamefont {Shiozaki},\ and\ \citenamefont
		{Yanase}}]{Sumita_2019}%
	\BibitemOpen
	\bibfield  {author} {\bibinfo {author} {\bibfnamefont {S.}~\bibnamefont
			{Sumita}}, \bibinfo {author} {\bibfnamefont {T.}~\bibnamefont {Nomoto}},
		\bibinfo {author} {\bibfnamefont {K.}~\bibnamefont {Shiozaki}},\ and\
		\bibinfo {author} {\bibfnamefont {Y.}~\bibnamefont {Yanase}},\ }\bibfield
	{title} {\bibinfo {title} {Classification of topological crystalline
			superconducting nodes on high-symmetry lines: Point nodes, line nodes, and
			bogoliubov fermi surfaces},\ }\href
	{https://doi.org/10.1103/PhysRevB.99.134513} {\bibfield  {journal} {\bibinfo
			{journal} {Phys. Rev. B}\ }\textbf {\bibinfo {volume} {99}},\ \bibinfo
		{pages} {134513} (\bibinfo {year} {2019})}\BibitemShut {NoStop}%
	\bibitem [{\citenamefont {Geier}\ \emph {et~al.}(2020)\citenamefont {Geier},
		\citenamefont {Brouwer},\ and\ \citenamefont {Trifunovic}}]{Geier_2020}%
	\BibitemOpen
	\bibfield  {author} {\bibinfo {author} {\bibfnamefont {M.}~\bibnamefont
			{Geier}}, \bibinfo {author} {\bibfnamefont {P.~W.}\ \bibnamefont {Brouwer}},\
		and\ \bibinfo {author} {\bibfnamefont {L.}~\bibnamefont {Trifunovic}},\
	}\bibfield  {title} {\bibinfo {title} {Symmetry-based indicators for
			topological bogoliubov--de gennes hamiltonians},\ }\href
	{https://doi.org/10.1103/PhysRevB.101.245128} {\bibfield  {journal} {\bibinfo
			{journal} {Phys. Rev. B}\ }\textbf {\bibinfo {volume} {101}},\ \bibinfo
		{pages} {245128} (\bibinfo {year} {2020})}\BibitemShut {NoStop}%
	\bibitem [{\citenamefont {Ono}\ \emph {et~al.}(2020)\citenamefont {Ono},
		\citenamefont {Po},\ and\ \citenamefont {Watanabe}}]{Ono_2020}%
	\BibitemOpen
	\bibfield  {author} {\bibinfo {author} {\bibfnamefont {S.}~\bibnamefont
			{Ono}}, \bibinfo {author} {\bibfnamefont {H.~C.}\ \bibnamefont {Po}},\ and\
		\bibinfo {author} {\bibfnamefont {H.}~\bibnamefont {Watanabe}},\ }\bibfield
	{title} {\bibinfo {title} {Refined symmetry indicators for topological
			superconductors in all space groups},\ }\href
	{https://doi.org/10.1126/sciadv.aaz8367} {\bibfield  {journal} {\bibinfo
			{journal} {Sci. Adv.}\ }\textbf {\bibinfo {volume} {6}},\ \bibinfo {pages}
		{eaaz8367} (\bibinfo {year} {2020})}\BibitemShut {NoStop}%
	\bibitem [{\citenamefont {Ueno}\ \emph {et~al.}(2013)\citenamefont {Ueno},
		\citenamefont {Yamakage}, \citenamefont {Tanaka},\ and\ \citenamefont
		{Sato}}]{Ueno_2013}%
	\BibitemOpen
	\bibfield  {author} {\bibinfo {author} {\bibfnamefont {Y.}~\bibnamefont
			{Ueno}}, \bibinfo {author} {\bibfnamefont {A.}~\bibnamefont {Yamakage}},
		\bibinfo {author} {\bibfnamefont {Y.}~\bibnamefont {Tanaka}},\ and\ \bibinfo
		{author} {\bibfnamefont {M.}~\bibnamefont {Sato}},\ }\bibfield  {title}
	{\bibinfo {title} {Symmetry-protected majorana fermions in topological
			crystalline superconductors: Theory and application to sr$_2$ruo$_4$},\
	}\href {https://doi.org/10.1103/PhysRevLett.111.087002} {\bibfield  {journal}
		{\bibinfo  {journal} {Phys. Rev. Lett.}\ }\textbf {\bibinfo {volume} {111}},\
		\bibinfo {pages} {087002} (\bibinfo {year} {2013})}\BibitemShut {NoStop}%
	\bibitem [{\citenamefont {Kobayashi}\ and\ \citenamefont
		{Furusaki}(2020)}]{Kobayashi_2020}%
	\BibitemOpen
	\bibfield  {author} {\bibinfo {author} {\bibfnamefont {S.}~\bibnamefont
			{Kobayashi}}\ and\ \bibinfo {author} {\bibfnamefont {A.}~\bibnamefont
			{Furusaki}},\ }\bibfield  {title} {\bibinfo {title} {Double majorana vortex
			zero modes in superconducting topological crystalline insulators with surface
			rotation anomaly},\ }\href {https://doi.org/10.1103/PhysRevB.102.180505}
	{\bibfield  {journal} {\bibinfo  {journal} {Phys. Rev. B}\ }\textbf {\bibinfo
			{volume} {102}},\ \bibinfo {pages} {180505(R)} (\bibinfo {year}
		{2020})}\BibitemShut {NoStop}%
	\bibitem [{\citenamefont {Zhang}\ \emph {et~al.}(2013)\citenamefont {Zhang},
		\citenamefont {Kane},\ and\ \citenamefont {Mele}}]{Zhang_2013}%
	\BibitemOpen
	\bibfield  {author} {\bibinfo {author} {\bibfnamefont {F.}~\bibnamefont
			{Zhang}}, \bibinfo {author} {\bibfnamefont {C.~L.}\ \bibnamefont {Kane}},\
		and\ \bibinfo {author} {\bibfnamefont {E.~J.}\ \bibnamefont {Mele}},\
	}\bibfield  {title} {\bibinfo {title} {Topological mirror
			superconductivity},\ }\href {https://doi.org/10.1103/PhysRevLett.111.056403}
	{\bibfield  {journal} {\bibinfo  {journal} {Phys. Rev. Lett.}\ }\textbf
		{\bibinfo {volume} {111}},\ \bibinfo {pages} {056403} (\bibinfo {year}
		{2013})}\BibitemShut {NoStop}%
	\bibitem [{\citenamefont {Chiu}\ and\ \citenamefont
		{Schnyder}(2014)}]{Chiu_2014}%
	\BibitemOpen
	\bibfield  {author} {\bibinfo {author} {\bibfnamefont {C.-K.}\ \bibnamefont
			{Chiu}}\ and\ \bibinfo {author} {\bibfnamefont {A.~P.}\ \bibnamefont
			{Schnyder}},\ }\bibfield  {title} {\bibinfo {title} {Classification of
			reflection-symmetry-protected topological semimetals and nodal
			superconductors},\ }\href {https://doi.org/10.1103/PhysRevB.90.205136}
	{\bibfield  {journal} {\bibinfo  {journal} {Phys. Rev. B}\ }\textbf {\bibinfo
			{volume} {90}},\ \bibinfo {pages} {205136} (\bibinfo {year}
		{2014})}\BibitemShut {NoStop}%
	\bibitem [{\citenamefont {Yan}\ \emph {et~al.}(2020)\citenamefont {Yan},
		\citenamefont {Wu},\ and\ \citenamefont {Huang}}]{Yan_2020}%
	\BibitemOpen
	\bibfield  {author} {\bibinfo {author} {\bibfnamefont {Z.}~\bibnamefont
			{Yan}}, \bibinfo {author} {\bibfnamefont {Z.}~\bibnamefont {Wu}},\ and\
		\bibinfo {author} {\bibfnamefont {W.}~\bibnamefont {Huang}},\ }\bibfield
	{title} {\bibinfo {title} {Vortex end majorana zero modes in superconducting
			dirac and weyl semimetals},\ }\href
	{https://doi.org/10.1103/PhysRevLett.124.257001} {\bibfield  {journal}
		{\bibinfo  {journal} {Phys. Rev. Lett.}\ }\textbf {\bibinfo {volume} {124}},\
		\bibinfo {pages} {257001} (\bibinfo {year} {2020})}\BibitemShut {NoStop}%
	\bibitem [{\citenamefont {Ghorashi}\ \emph {et~al.}(2020)\citenamefont
		{Ghorashi}, \citenamefont {Hughes},\ and\ \citenamefont
		{Rossi}}]{Ghorashi_2020}%
	\BibitemOpen
	\bibfield  {author} {\bibinfo {author} {\bibfnamefont {S.~A.~A.}\
			\bibnamefont {Ghorashi}}, \bibinfo {author} {\bibfnamefont {T.~L.}\
			\bibnamefont {Hughes}},\ and\ \bibinfo {author} {\bibfnamefont
			{E.}~\bibnamefont {Rossi}},\ }\bibfield  {title} {\bibinfo {title} {Vortex
			and surface phase transitions in superconducting higher-order topological
			insulators},\ }\href {https://doi.org/10.1103/PhysRevLett.125.037001}
	{\bibfield  {journal} {\bibinfo  {journal} {Phys. Rev. Lett.}\ }\textbf
		{\bibinfo {volume} {125}},\ \bibinfo {pages} {037001} (\bibinfo {year}
		{2020})}\BibitemShut {NoStop}%
	\bibitem [{\citenamefont {Hosur}\ \emph {et~al.}(2011)\citenamefont {Hosur},
		\citenamefont {Ghaemi}, \citenamefont {Mong},\ and\ \citenamefont
		{Vishwanath}}]{Hosur_2011}%
	\BibitemOpen
	\bibfield  {author} {\bibinfo {author} {\bibfnamefont {P.}~\bibnamefont
			{Hosur}}, \bibinfo {author} {\bibfnamefont {P.}~\bibnamefont {Ghaemi}},
		\bibinfo {author} {\bibfnamefont {R.~S.~K.}\ \bibnamefont {Mong}},\ and\
		\bibinfo {author} {\bibfnamefont {A.}~\bibnamefont {Vishwanath}},\ }\bibfield
	{title} {\bibinfo {title} {Majorana modes at the ends of superconductor
			vortices in doped topological insulators},\ }\href
	{https://doi.org/10.1103/PhysRevLett.107.097001} {\bibfield  {journal}
		{\bibinfo  {journal} {Phys. Rev. Lett.}\ }\textbf {\bibinfo {volume} {107}},\
		\bibinfo {pages} {097001} (\bibinfo {year} {2011})}\BibitemShut {NoStop}%
	\bibitem [{\citenamefont {Kheirkhah}\ \emph {et~al.}(2021)\citenamefont
		{Kheirkhah}, \citenamefont {Yan},\ and\ \citenamefont
		{Marsiglio}}]{Kheirkhah2020}%
	\BibitemOpen
	\bibfield  {author} {\bibinfo {author} {\bibfnamefont {M.}~\bibnamefont
			{Kheirkhah}}, \bibinfo {author} {\bibfnamefont {Z.}~\bibnamefont {Yan}},\
		and\ \bibinfo {author} {\bibfnamefont {F.}~\bibnamefont {Marsiglio}},\
	}\bibfield  {title} {\bibinfo {title} {Vortex-line topology in iron-based
			superconductors with and without second-order topology},\ }\href
	{https://doi.org/10.1103/PhysRevB.103.L140502} {\bibfield  {journal}
		{\bibinfo  {journal} {Phys. Rev. B}\ }\textbf {\bibinfo {volume} {103}},\
		\bibinfo {pages} {L140502} (\bibinfo {year} {2021})}\BibitemShut {NoStop}%
	\bibitem [{\citenamefont {Qin}\ \emph {et~al.}(2019)\citenamefont {Qin},
		\citenamefont {Hu}, \citenamefont {Le}, \citenamefont {Zeng}, \citenamefont
		{Zhang}, \citenamefont {Fang},\ and\ \citenamefont {Hu}}]{Qin_2019}%
	\BibitemOpen
	\bibfield  {author} {\bibinfo {author} {\bibfnamefont {S.}~\bibnamefont
			{Qin}}, \bibinfo {author} {\bibfnamefont {L.}~\bibnamefont {Hu}}, \bibinfo
		{author} {\bibfnamefont {C.}~\bibnamefont {Le}}, \bibinfo {author}
		{\bibfnamefont {J.}~\bibnamefont {Zeng}}, \bibinfo {author} {\bibfnamefont
			{F.-c.}\ \bibnamefont {Zhang}}, \bibinfo {author} {\bibfnamefont
			{C.}~\bibnamefont {Fang}},\ and\ \bibinfo {author} {\bibfnamefont
			{J.}~\bibnamefont {Hu}},\ }\bibfield  {title} {\bibinfo {title} {Quasi-1d
			topological nodal vortex line phase in doped superconducting 3d dirac
			semimetals},\ }\href {https://doi.org/10.1103/PhysRevLett.123.027003}
	{\bibfield  {journal} {\bibinfo  {journal} {Phys. Rev. Lett.}\ }\textbf
		{\bibinfo {volume} {123}},\ \bibinfo {pages} {027003} (\bibinfo {year}
		{2019})}\BibitemShut {NoStop}%
	\bibitem [{\citenamefont {Kitaev}\ \emph {et~al.}(2009)\citenamefont {Kitaev},
		\citenamefont {Lebedev},\ and\ \citenamefont {Feigel'man}}]{Kitaev_2009}%
	\BibitemOpen
	\bibfield  {author} {\bibinfo {author} {\bibfnamefont {A.}~\bibnamefont
			{Kitaev}}, \bibinfo {author} {\bibfnamefont {V.}~\bibnamefont {Lebedev}},\
		and\ \bibinfo {author} {\bibfnamefont {M.}~\bibnamefont {Feigel'man}},\
	}\bibfield  {title} {\bibinfo {title} {Periodic table for topological
			insulators and superconductors},\ }\href {https://doi.org/10.1063/1.3149495}
	{\bibfield  {journal} {\bibinfo  {journal} {AIP Conf. Proc.}\ }\textbf
		{\bibinfo {volume} {1134}},\ \bibinfo {pages} {22} (\bibinfo {year}
		{2009})}\BibitemShut {NoStop}%
	\bibitem [{\citenamefont {Ryu}\ \emph {et~al.}(2010)\citenamefont {Ryu},
		\citenamefont {Schnyder}, \citenamefont {Furusaki},\ and\ \citenamefont
		{Ludwig}}]{Ryu_2010}%
	\BibitemOpen
	\bibfield  {author} {\bibinfo {author} {\bibfnamefont {S.}~\bibnamefont
			{Ryu}}, \bibinfo {author} {\bibfnamefont {A.~P.}\ \bibnamefont {Schnyder}},
		\bibinfo {author} {\bibfnamefont {A.}~\bibnamefont {Furusaki}},\ and\
		\bibinfo {author} {\bibfnamefont {A.~W.~W.}\ \bibnamefont {Ludwig}},\
	}\bibfield  {title} {\bibinfo {title} {Topological insulators and
			superconductors: tenfold way and dimensional hierarchy},\ }\href
	{https://doi.org/10.1088/1367-2630/12/6/065010} {\bibfield  {journal}
		{\bibinfo  {journal} {New J. Phys.}\ }\textbf {\bibinfo {volume} {12}},\
		\bibinfo {pages} {065010} (\bibinfo {year} {2010})}\BibitemShut {NoStop}%
	\bibitem [{sup()}]{supp}%
	\BibitemOpen
	\href@noop {} {\bibinfo {title} {See supplemental material for fragile
			topology in nodal-line semimetal superconductors.}}\BibitemShut {Stop}%
	\bibitem [{\citenamefont {Alexandradinata}\ \emph {et~al.}(2016)\citenamefont
		{Alexandradinata}, \citenamefont {Wang},\ and\ \citenamefont
		{Bernevig}}]{Alexandradinata_2016}%
	\BibitemOpen
	\bibfield  {author} {\bibinfo {author} {\bibfnamefont {A.}~\bibnamefont
			{Alexandradinata}}, \bibinfo {author} {\bibfnamefont {Z.}~\bibnamefont
			{Wang}},\ and\ \bibinfo {author} {\bibfnamefont {B.~A.}\ \bibnamefont
			{Bernevig}},\ }\bibfield  {title} {\bibinfo {title} {Topological insulators
			from group cohomology},\ }\href {https://doi.org/10.1103/PhysRevX.6.021008}
	{\bibfield  {journal} {\bibinfo  {journal} {Phys. Rev. X}\ }\textbf {\bibinfo
			{volume} {6}},\ \bibinfo {pages} {021008} (\bibinfo {year}
		{2016})}\BibitemShut {NoStop}%
	\bibitem [{\citenamefont {Wieder}\ \emph {et~al.}(2018)\citenamefont {Wieder},
		\citenamefont {Bradlyn}, \citenamefont {Wang}, \citenamefont {Cano},
		\citenamefont {Kim}, \citenamefont {Kim}, \citenamefont {Rappe},
		\citenamefont {Kane},\ and\ \citenamefont {Bernevig}}]{Wieder_2018}%
	\BibitemOpen
	\bibfield  {author} {\bibinfo {author} {\bibfnamefont {B.~J.}\ \bibnamefont
			{Wieder}}, \bibinfo {author} {\bibfnamefont {B.}~\bibnamefont {Bradlyn}},
		\bibinfo {author} {\bibfnamefont {Z.}~\bibnamefont {Wang}}, \bibinfo {author}
		{\bibfnamefont {J.}~\bibnamefont {Cano}}, \bibinfo {author} {\bibfnamefont
			{Y.}~\bibnamefont {Kim}}, \bibinfo {author} {\bibfnamefont {H.-S.~D.}\
			\bibnamefont {Kim}}, \bibinfo {author} {\bibfnamefont {A.~M.}\ \bibnamefont
			{Rappe}}, \bibinfo {author} {\bibfnamefont {C.~L.}\ \bibnamefont {Kane}},\
		and\ \bibinfo {author} {\bibfnamefont {B.~A.}\ \bibnamefont {Bernevig}},\
	}\bibfield  {title} {\bibinfo {title} {Wallpaper fermions and the
			nonsymmorphic dirac insulator},\ }\href
	{https://doi.org/10.1126/science.aan2802} {\bibfield  {journal} {\bibinfo
			{journal} {Science}\ }\textbf {\bibinfo {volume} {361}},\ \bibinfo {pages}
		{246} (\bibinfo {year} {2018})}\BibitemShut {NoStop}%
	\bibitem [{\citenamefont {Alexandradinata}\ \emph {et~al.}(2014)\citenamefont
		{Alexandradinata}, \citenamefont {Dai},\ and\ \citenamefont
		{Bernevig}}]{Alexandradinata_2014}%
	\BibitemOpen
	\bibfield  {author} {\bibinfo {author} {\bibfnamefont {A.}~\bibnamefont
			{Alexandradinata}}, \bibinfo {author} {\bibfnamefont {X.}~\bibnamefont
			{Dai}},\ and\ \bibinfo {author} {\bibfnamefont {B.~A.}\ \bibnamefont
			{Bernevig}},\ }\bibfield  {title} {\bibinfo {title} {Wilson-loop
			characterization of inversion-symmetric topological insulators},\ }\href
	{https://doi.org/10.1103/PhysRevB.89.155114} {\bibfield  {journal} {\bibinfo
			{journal} {Phys. Rev. B}\ }\textbf {\bibinfo {volume} {89}},\ \bibinfo
		{pages} {155114} (\bibinfo {year} {2014})}\BibitemShut {NoStop}%
	\bibitem [{\citenamefont {Fu}\ and\ \citenamefont {Kane}(2007)}]{Fu_2007a}%
	\BibitemOpen
	\bibfield  {author} {\bibinfo {author} {\bibfnamefont {L.}~\bibnamefont
			{Fu}}\ and\ \bibinfo {author} {\bibfnamefont {C.~L.}\ \bibnamefont {Kane}},\
	}\bibfield  {title} {\bibinfo {title} {Topological insulators with inversion
			symmetry},\ }\href {https://doi.org/10.1103/PhysRevB.76.045302} {\bibfield
		{journal} {\bibinfo  {journal} {Phys. Rev. B}\ }\textbf {\bibinfo {volume}
			{76}},\ \bibinfo {pages} {045302} (\bibinfo {year} {2007})}\BibitemShut
	{NoStop}%
	\bibitem [{\citenamefont {Kitaev}(2001)}]{Kitaev_2001}%
	\BibitemOpen
	\bibfield  {author} {\bibinfo {author} {\bibfnamefont {A.~Y.}\ \bibnamefont
			{Kitaev}},\ }\bibfield  {title} {\bibinfo {title} {Unpaired majorana fermions
			in quantum wires},\ }\href {https://doi.org/10.1070/1063-7869/44/10S/S29}
	{\bibfield  {journal} {\bibinfo  {journal} {Physics-Uspekhi}\ }\textbf
		{\bibinfo {volume} {44}},\ \bibinfo {pages} {131} (\bibinfo {year}
		{2001})}\BibitemShut {NoStop}%
	\bibitem [{\citenamefont {Budich}\ and\ \citenamefont
		{Ardonne}(2013)}]{Budich_2013}%
	\BibitemOpen
	\bibfield  {author} {\bibinfo {author} {\bibfnamefont {J.~C.}\ \bibnamefont
			{Budich}}\ and\ \bibinfo {author} {\bibfnamefont {E.}~\bibnamefont
			{Ardonne}},\ }\bibfield  {title} {\bibinfo {title} {Equivalent topological
			invariants for one-dimensional majorana wires in symmetry {classD}},\ }\href
	{https://doi.org/10.1103/PhysRevB.88.075419} {\bibfield  {journal} {\bibinfo
			{journal} {Phys. Rev. B}\ }\textbf {\bibinfo {volume} {88}},\ \bibinfo
		{pages} {075419} (\bibinfo {year} {2013})}\BibitemShut {NoStop}%
	\bibitem [{not()}]{note}%
	\BibitemOpen
	\href@noop {} {\bibinfo {title} {We have checked numerically that our main
			results for the vortex line effect in both $s$-wave and $p$-wave nlsm
			superconductors are robust considering different non-zero chemical potentials
			with $0.1\leq\mu\leq0.4$.}}\BibitemShut {Stop}%
	\bibitem [{\citenamefont {Ghosh}\ \emph {et~al.}(2021)\citenamefont {Ghosh},
		\citenamefont {Nag},\ and\ \citenamefont {Saha}}]{PhysRevB.104.134508}%
	\BibitemOpen
	\bibfield  {author} {\bibinfo {author} {\bibfnamefont {A.~K.}\ \bibnamefont
			{Ghosh}}, \bibinfo {author} {\bibfnamefont {T.}~\bibnamefont {Nag}},\ and\
		\bibinfo {author} {\bibfnamefont {A.}~\bibnamefont {Saha}},\ }\bibfield
	{title} {\bibinfo {title} {Hierarchy of higher-order topological
			superconductors in three dimensions},\ }\href
	{https://doi.org/10.1103/PhysRevB.104.134508} {\bibfield  {journal} {\bibinfo
			{journal} {Phys. Rev. B}\ }\textbf {\bibinfo {volume} {104}},\ \bibinfo
		{pages} {134508} (\bibinfo {year} {2021})}\BibitemShut {NoStop}%
	\bibitem [{\citenamefont {Hu}\ \emph {et~al.}(2021)\citenamefont {Hu},
		\citenamefont {Wu}, \citenamefont {Liu},\ and\ \citenamefont
		{Zhang}}]{hu2021competing}%
	\BibitemOpen
	\bibfield  {author} {\bibinfo {author} {\bibfnamefont {L.-H.}\ \bibnamefont
			{Hu}}, \bibinfo {author} {\bibfnamefont {X.}~\bibnamefont {Wu}}, \bibinfo
		{author} {\bibfnamefont {C.-X.}\ \bibnamefont {Liu}},\ and\ \bibinfo {author}
		{\bibfnamefont {R.-X.}\ \bibnamefont {Zhang}},\ }\href@noop {} {\bibinfo
		{title} {Competing vortex topologies in iron-based superconductors}}
	(\bibinfo {year} {2021}),\ \Eprint {https://arxiv.org/abs/2110.11357}
	{arXiv:2110.11357 [cond-mat.supr-con]} \BibitemShut {NoStop}%
\end{thebibliography}

\begin{thebibliography}{9}%
	\makeatletter
	\providecommand \@ifxundefined [1]{%
		\@ifx{#1\undefined}
	}%
	\providecommand \@ifnum [1]{%
		\ifnum #1\expandafter \@firstoftwo
		\else \expandafter \@secondoftwo
		\fi
	}%
	\providecommand \@ifx [1]{%
		\ifx #1\expandafter \@firstoftwo
		\else \expandafter \@secondoftwo
		\fi
	}%
	\providecommand \natexlab [1]{#1}%
	\providecommand \enquote  [1]{``#1''}%
	\providecommand \bibnamefont  [1]{#1}%
	\providecommand \bibfnamefont [1]{#1}%
	\providecommand \citenamefont [1]{#1}%
	\providecommand \href@noop [0]{\@secondoftwo}%
	\providecommand \href [0]{\begingroup \@sanitize@url \@href}%
	\providecommand \@href[1]{\@@startlink{#1}\@@href}%
	\providecommand \@@href[1]{\endgroup#1\@@endlink}%
	\providecommand \@sanitize@url [0]{\catcode `\\12\catcode `\$12\catcode
		`\&12\catcode `\#12\catcode `\^12\catcode `\_12\catcode `\%12\relax}%
	\providecommand \@@startlink[1]{}%
	\providecommand \@@endlink[0]{}%
	\providecommand \url  [0]{\begingroup\@sanitize@url \@url }%
	\providecommand \@url [1]{\endgroup\@href {#1}{\urlprefix }}%
	\providecommand \urlprefix  [0]{URL }%
	\providecommand \Eprint [0]{\href }%
	\providecommand \doibase [0]{https://doi.org/}%
	\providecommand \selectlanguage [0]{\@gobble}%
	\providecommand \bibinfo  [0]{\@secondoftwo}%
	\providecommand \bibfield  [0]{\@secondoftwo}%
	\providecommand \translation [1]{[#1]}%
	\providecommand \BibitemOpen [0]{}%
	\providecommand \bibitemStop [0]{}%
	\providecommand \bibitemNoStop [0]{.\EOS\space}%
	\providecommand \EOS [0]{\spacefactor3000\relax}%
	\providecommand \BibitemShut  [1]{\csname bibitem#1\endcsname}%
	\let\auto@bib@innerbib\@empty
	\bibitem [{\citenamefont {Alexandradinata}\ \emph {et~al.}(2016)\citenamefont
		{Alexandradinata}, \citenamefont {Wang},\ and\ \citenamefont
		{Bernevig}}]{Alexandradinata_2016}%
	\BibitemOpen
	\bibfield  {author} {\bibinfo {author} {\bibfnamefont {A.}~\bibnamefont
			{Alexandradinata}}, \bibinfo {author} {\bibfnamefont {Z.}~\bibnamefont
			{Wang}},\ and\ \bibinfo {author} {\bibfnamefont {B.~A.}\ \bibnamefont
			{Bernevig}},\ }\bibfield  {title} {\bibinfo {title} {Topological insulators
			from group cohomology},\ }\href {https://doi.org/10.1103/PhysRevX.6.021008}
	{\bibfield  {journal} {\bibinfo  {journal} {Phys. Rev. X}\ }\textbf {\bibinfo
			{volume} {6}},\ \bibinfo {pages} {021008} (\bibinfo {year}
		{2016})}\BibitemShut {NoStop}%
	\bibitem [{\citenamefont {Wieder}\ \emph {et~al.}(2018)\citenamefont {Wieder},
		\citenamefont {Bradlyn}, \citenamefont {Wang}, \citenamefont {Cano},
		\citenamefont {Kim}, \citenamefont {Kim}, \citenamefont {Rappe},
		\citenamefont {Kane},\ and\ \citenamefont {Bernevig}}]{Wieder_2018}%
	\BibitemOpen
	\bibfield  {author} {\bibinfo {author} {\bibfnamefont {B.~J.}\ \bibnamefont
			{Wieder}}, \bibinfo {author} {\bibfnamefont {B.}~\bibnamefont {Bradlyn}},
		\bibinfo {author} {\bibfnamefont {Z.}~\bibnamefont {Wang}}, \bibinfo {author}
		{\bibfnamefont {J.}~\bibnamefont {Cano}}, \bibinfo {author} {\bibfnamefont
			{Y.}~\bibnamefont {Kim}}, \bibinfo {author} {\bibfnamefont {H.-S.~D.}\
			\bibnamefont {Kim}}, \bibinfo {author} {\bibfnamefont {A.~M.}\ \bibnamefont
			{Rappe}}, \bibinfo {author} {\bibfnamefont {C.~L.}\ \bibnamefont {Kane}},\
		and\ \bibinfo {author} {\bibfnamefont {B.~A.}\ \bibnamefont {Bernevig}},\
	}\bibfield  {title} {\bibinfo {title} {Wallpaper fermions and the
			nonsymmorphic dirac insulator},\ }\href
	{https://doi.org/10.1126/science.aan2802} {\bibfield  {journal} {\bibinfo
			{journal} {Science}\ }\textbf {\bibinfo {volume} {361}},\ \bibinfo {pages}
		{246} (\bibinfo {year} {2018})}\BibitemShut {NoStop}%
	\bibitem [{\citenamefont {Alexandradinata}\ \emph {et~al.}(2014)\citenamefont
		{Alexandradinata}, \citenamefont {Dai},\ and\ \citenamefont
		{Bernevig}}]{Alexandradinata_2014}%
	\BibitemOpen
	\bibfield  {author} {\bibinfo {author} {\bibfnamefont {A.}~\bibnamefont
			{Alexandradinata}}, \bibinfo {author} {\bibfnamefont {X.}~\bibnamefont
			{Dai}},\ and\ \bibinfo {author} {\bibfnamefont {B.~A.}\ \bibnamefont
			{Bernevig}},\ }\bibfield  {title} {\bibinfo {title} {Wilson-loop
			characterization of inversion-symmetric topological insulators},\ }\href
	{https://doi.org/10.1103/PhysRevB.89.155114} {\bibfield  {journal} {\bibinfo
			{journal} {Phys. Rev. B}\ }\textbf {\bibinfo {volume} {89}},\ \bibinfo
		{pages} {155114} (\bibinfo {year} {2014})}\BibitemShut {NoStop}%
	\bibitem [{\citenamefont {Ghosh}\ \emph {et~al.}(2021)\citenamefont {Ghosh},
		\citenamefont {Nag},\ and\ \citenamefont {Saha}}]{PhysRevB.104.134508}%
	\BibitemOpen
	\bibfield  {author} {\bibinfo {author} {\bibfnamefont {A.~K.}\ \bibnamefont
			{Ghosh}}, \bibinfo {author} {\bibfnamefont {T.}~\bibnamefont {Nag}},\ and\
		\bibinfo {author} {\bibfnamefont {A.}~\bibnamefont {Saha}},\ }\bibfield
	{title} {\bibinfo {title} {Hierarchy of higher-order topological
			superconductors in three dimensions},\ }\href
	{https://doi.org/10.1103/PhysRevB.104.134508} {\bibfield  {journal} {\bibinfo
			{journal} {Phys. Rev. B}\ }\textbf {\bibinfo {volume} {104}},\ \bibinfo
		{pages} {134508} (\bibinfo {year} {2021})}\BibitemShut {NoStop}%
	\bibitem [{\citenamefont {Benalcazar}\ \emph {et~al.}(2017)\citenamefont
		{Benalcazar}, \citenamefont {Bernevig},\ and\ \citenamefont
		{Hughes}}]{PhysRevB.96.245115}%
	\BibitemOpen
	\bibfield  {author} {\bibinfo {author} {\bibfnamefont {W.~A.}\ \bibnamefont
			{Benalcazar}}, \bibinfo {author} {\bibfnamefont {B.~A.}\ \bibnamefont
			{Bernevig}},\ and\ \bibinfo {author} {\bibfnamefont {T.~L.}\ \bibnamefont
			{Hughes}},\ }\bibfield  {title} {\bibinfo {title} {Electric multipole
			moments, topological multipole moment pumping, and chiral hinge states in
			crystalline insulators},\ }\href {https://doi.org/10.1103/PhysRevB.96.245115}
	{\bibfield  {journal} {\bibinfo  {journal} {Phys. Rev. B}\ }\textbf {\bibinfo
			{volume} {96}},\ \bibinfo {pages} {245115} (\bibinfo {year}
		{2017})}\BibitemShut {NoStop}%
	\bibitem [{\citenamefont {Fu}\ and\ \citenamefont {Kane}(2007)}]{Fu_2007a}%
	\BibitemOpen
	\bibfield  {author} {\bibinfo {author} {\bibfnamefont {L.}~\bibnamefont
			{Fu}}\ and\ \bibinfo {author} {\bibfnamefont {C.~L.}\ \bibnamefont {Kane}},\
	}\bibfield  {title} {\bibinfo {title} {Topological insulators with inversion
			symmetry},\ }\href {https://doi.org/10.1103/PhysRevB.76.045302} {\bibfield
		{journal} {\bibinfo  {journal} {Phys. Rev. B}\ }\textbf {\bibinfo {volume}
			{76}},\ \bibinfo {pages} {045302} (\bibinfo {year} {2007})}\BibitemShut
	{NoStop}%
	\bibitem [{\citenamefont {Kitaev}(2001)}]{Kitaev_2001}%
	\BibitemOpen
	\bibfield  {author} {\bibinfo {author} {\bibfnamefont {A.~Y.}\ \bibnamefont
			{Kitaev}},\ }\bibfield  {title} {\bibinfo {title} {Unpaired majorana fermions
			in quantum wires},\ }\href {https://doi.org/10.1070/1063-7869/44/10S/S29}
	{\bibfield  {journal} {\bibinfo  {journal} {Physics-Uspekhi}\ }\textbf
		{\bibinfo {volume} {44}},\ \bibinfo {pages} {131} (\bibinfo {year}
		{2001})}\BibitemShut {NoStop}%
	\bibitem [{\citenamefont {Budich}\ and\ \citenamefont
		{Ardonne}(2013)}]{Budich_2013}%
	\BibitemOpen
	\bibfield  {author} {\bibinfo {author} {\bibfnamefont {J.~C.}\ \bibnamefont
			{Budich}}\ and\ \bibinfo {author} {\bibfnamefont {E.}~\bibnamefont
			{Ardonne}},\ }\bibfield  {title} {\bibinfo {title} {Equivalent topological
			invariants for one-dimensional majorana wires in symmetry {classD}},\ }\href
	{https://doi.org/10.1103/PhysRevB.88.075419} {\bibfield  {journal} {\bibinfo
			{journal} {Phys. Rev. B}\ }\textbf {\bibinfo {volume} {88}},\ \bibinfo
		{pages} {075419} (\bibinfo {year} {2013})}\BibitemShut {NoStop}%
	\bibitem [{\citenamefont {Wimmer}(2012)}]{Wimmer_2012}%
	\BibitemOpen
	\bibfield  {author} {\bibinfo {author} {\bibfnamefont {M.}~\bibnamefont
			{Wimmer}},\ }\bibfield  {title} {\bibinfo {title} {Algorithm 923},\ }\href
	{https://doi.org/10.1145/2331130.2331138} {\bibfield  {journal} {\bibinfo
			{journal} {ACM Trans. Math. Softw.}\ }\textbf {\bibinfo {volume} {38}},\
		\bibinfo {pages} {1} (\bibinfo {year} {2012})}\BibitemShut {NoStop}%
\end{thebibliography}

%

\renewcommand{\thesection}{S-\arabic{section}}
\setcounter{section}{0}  
\renewcommand{\theequation}{S\arabic{equation}}
\setcounter{equation}{0}  
\renewcommand{\thefigure}{S\arabic{figure}}
\setcounter{figure}{0}  
\renewcommand{\thetable}{S\Roman{table}}
\setcounter{table}{0}

\onecolumngrid
\flushbottom

\begin{center}\large
	\textbf{Supplementary Material}
\end{center}
\flushbottom

\subsection{The Winding number}
1. The winding number

Considering a closed loop for the integral circling around the path from $(k_{x},k_{y},-\pi)$ to $(k_{x},k_{y},\pi)$, the relative phase between any two states is expressed as:\\
\begin{eqnarray}
	\exp(-i\theta_{n})&&=\left\langle u^{n}(k_{x},k_{y},k_{z}) \right| \left.  u^{n}(k_{x},k_{y},k_{z}+\Delta k_{z})\right\rangle,
\end{eqnarray}
where $\left|  u^{n}(k_{x},k_{y},k_{z}) \right\rangle  $ is the eigenvector of $H^{s/p}$ corresponding to the occupied band.\\

Dividing the closed path into $N$ segments and adding up the relative phases, we have
\begin{eqnarray}
	{\mathrm M}_{n}=&&\langle u^{n}(k_{x},k_{y},-\pi) |   u^{n}(k_{x},k_{y},-\pi+\Delta k_{z})\rangle \langle u^{n}(k_{x},k_{y},-\pi+\Delta k_{z}) |   u^{n}(k_{x},k_{y},-\pi+2\Delta k_{z})\rangle \nonumber \\
	&&\dots \langle u^{n}(k_{x},k_{y},-\pi+(N-1)\Delta k_{z}) | \  u^{n}(k_{x},k_{y},\pi)\rangle,
\end{eqnarray}
with $\Delta k_z=2\pi/N$.

The winding number is then calculated as:
\begin{eqnarray}
	{\mathrm w}_{n}&&=-{\mathrm{Im}} \quad \ln ({\mathrm M}_{n})/\pi
\end{eqnarray}

\subsection{The Wilson loop method}

The Wilson loop is calculated based the method introduced in Refs.~\cite{Alexandradinata_2016,Wieder_2018,Alexandradinata_2014}.
Considering the winding direction along the $k_z$ direction, we define a Wilson loop matrix with the elements being written as,

\begin{eqnarray}
	[W(k_{x},k_{y})]_{ij}=&&\langle u^{i}(k_{x},k_{y},-\pi) | P^{1}(k_{x},k_{y},-\pi+\Delta k_{z})  P^{2}(k_{x},k_{y},-\pi+2\Delta k_{z}) \nonumber \\
	&&\dots  P^{N}(k_{x},k_{y},-\pi+(N-1)\Delta k_{z}) |   u^{j}(k_{x},k_{y},\pi)\rangle.
\end{eqnarray}
$P^{m}(k_x,k_y,K_z+m\Delta k_z)$ is the projection operator with $P^{m}({\bf k})=\sum_n |u^n({\bf k})\rangle\langle u^n({\bf k})$.

Then we can define the Wilson phase ${\mathrm \theta(k_{x},k_{y})}$ to describe the Wilson loop spectrum, which can be calculated from diagonalizing the Wilson loop matrix, with
\begin{eqnarray}
	{\mathrm \theta(k_{x},k_{y})}&&={\mathrm{arg}}[\mathrm{eig}(W(k_{x},k_{y}))],
\end{eqnarray}
with $\mathrm{eig}(W)$ being the eigenvalues of the matrix $W$. $\mathrm{arg}(A)$ is the argument of the complex number $A$.\\

\subsection{The Wannier spectra}

We study the Wannier spectra according to Refs.~\cite{PhysRevB.104.134508,PhysRevB.96.245115}.
We first define a matrix with the elements being expressed as,
\begin{eqnarray}
	[W^{x}(y,k_{z}=0)]_{ij}=M^{x}_{k_{y}+(N-1)\triangle k_{y}}\dots M^{x}_{k_{y}+\Delta k_{y}}M^{x}_{k_{y}},
\end{eqnarray}
where $[M^{x}_{k_{y}}]_{ij}=\left\langle u^{x}_{j,k_{y}+\triangle k_{y}}\right| \left.u^{x}_{i,k_{y}}\right\rangle $. $\Delta k_{y}=2\pi/N$. $\left| u^{x}_{i,k_{y}}\right\rangle $ are the occupied state of the spinful $s$-wave SC-NLSM Hamiltonian with considering open boundary condition along the $x$-direction and periodic boundary condition in the $y-z$ plane.
The corresponding Wannier Hamiltonian is calculated to be $H_{W^{x}(y,k_{z}=0)}=-i\ln W^{x}(y,k_{z}=0)$. The wannier spectra $v^{x}_{y,k_{z}=0}$ is then expressed as,

\begin{eqnarray}
	v^{x}_{y,k_{z}=0}=\mathrm{eig}(H_{W^{x}(y,k_{z}=0)})/(2\pi).
\end{eqnarray}

\subsection{${\mathrm Z_{2}}$ topological invariant of spinless SC-NLSM with the vortex line}
Considering the open boundary condition along the $x$ and the $y$ directions and the periodic boundary condition along the $z$ direction with a vortex at the center of the $x-y$ plane, the Hamiltonian is expressed as,
\begin{eqnarray}
	H^{s}(k_{z})=&&\sum_{{\bf \hat{r}, \sigma}}[-\mu c^{\dagger}_{{\bf \hat{r}},\sigma}\sigma_{0}c_{{\bf \hat{r}},\sigma} + (m-6+2t_{z}\cos(k_z))c^{\dagger}_{{\bf \hat{r}},\sigma}\sigma_{x}c_{{\bf \hat{r}},\sigma}+t_{x}c^{\dagger}_{{\bf \hat{r}},\sigma}\sigma_{x}c_{{\bf \hat{r}+\hat{x}},\sigma} \nonumber \\
	&&+t_{y}c^{\dagger}_{{\bf \hat{r}},\sigma}\sigma_{x}c_{{\bf \hat{r}+\hat{y}},\sigma}+\Delta_{s}({\bf \hat{r}})c^{\dagger}_{{\bf \hat{r}},\sigma}(-i\sigma_{y})c^{\dagger}_{{\bf \hat{r}},\sigma}+\lambda_{z}c^{\dagger}_{{\bf \hat{r}},\sigma}\sigma_{z}c_{{\bf \hat{r}},\sigma}],
\end{eqnarray}
and
\begin{eqnarray}
	H^{p}(k_{z})=&&\sum_{{\bf \hat{r}, \sigma}}[-\mu c^{\dagger}_{{\bf \hat{r}},\sigma} \sigma_{0}c_{{\bf \hat{r}},\sigma} + (m-6+2t_{z}\cos(kz))c^{\dagger}_{{\bf \hat{r}},\sigma}\sigma_{x}c_{{\bf \hat{r}},\sigma}+t_{x}c^{\dagger}_{{\bf \hat{r}},\sigma}\sigma_{x}c_{{\bf \hat{r}+\hat{x}},\sigma} \nonumber \\
	&&+t_{y}c^{\dagger}_{{\bf \hat{r}},\sigma}\sigma_{x}c_{{\bf \hat{r}+\hat{y}},\sigma}-i\Delta_{p}({\bf \hat{r}})c^{\dagger}_{{\bf \hat{r}},\sigma}\sigma_{0}c^{\dagger}_{{\bf \hat{r}+\hat{x}},\sigma}+\Delta_{p}({\bf \hat{r}})c^{\dagger}_{{\bf \hat{r}},\sigma}\sigma_{0}c^{\dagger}_{{\bf \hat{r}+\hat{y}},\sigma}+\lambda_{z}c^{\dagger}_{{\bf \hat{r}},\sigma}\sigma_{z}c_{{\bf \hat{r}},\sigma}].
\end{eqnarray}
${\bf \hat{r}}$ and $\sigma$ represent the coordinate of the lattice sites and the orbital, respectively. The superconducting order parameter with a vortex line in the real space is expressed as,
\begin{eqnarray}
	\Delta_{s}({\bf \hat{r}})&&=\Delta_{0}\tanh(\sqrt{x_{i}^{2}+y_{i}^{2}}/\zeta)e^{i\theta} \\
	\Delta_{p}({\bf \hat{r}})&&=\Delta_{p}\tanh(\sqrt{(\dfrac{x_{i}+x_{j}}{2})^{2}+(\dfrac{y_{i}+y_{j}}{2})^{2}}/\zeta)e^{i\theta},
\end{eqnarray}
where $(x_{i},y_{i})$ and $(x_{j},y_{j})$ are two nearest-neighbour sites, respectively.
With the vortex line, the inversion symmetry and the mirror symmetry of SC-NLSM still preserve, with the inversion operation and the mirror operation being expressed as,
$(x,y,z)\rightarrow(-x,-y,-z)$ and $(x,y,z)\rightarrow(x,y,-z)$, respectively.
The particle and hole parts of Eq.(S8) and Eq.(S9) is inversion and mirror symmetric. Let us analysis the pairing term with a vortex line. The spinless $s$-wave superconducting term is express as,
\begin{eqnarray}
	H^s_{sc}=\sum_{k_{z},x,y}[e^{i\theta(x,y)}\Delta_{s}(k_{z},x,y)c_{a}^{\dagger}(k_{z},x,y)c_{b}^{\dagger}(-k_{z},x,y)+h.c],
\end{eqnarray}
where $\theta(x,y)=-\theta(-x,-y)$.\\
Under the inversion symmetry operation, we have
\begin{eqnarray}
	IH^s_{sc}I^{-1}=-\sum_{k_{z},x,y}[e^{i\theta(-x,-y)}\Delta_{s}(-k_{z},-x,-y)c_{b}^{\dagger}(-k_{z},-x,-y)c_{a}^{\dagger}(k_{z},-x,-y)+h.c]=H^s_{sc}.
\end{eqnarray}
Under the mirror symmetry operation, we have
\begin{eqnarray}
	M_{xy}H^s_{sc}M_{xy}^{-1}=\sum_{k_{z},x,y}[e^{i\theta(x,y)}\Delta_{s}(-k_{z},x,y)c_{b}^{\dagger}(-k_{z},x,y)c_{a}^{\dagger}(k_{z},x,y)+h.c]=H^s_{sc}.
\end{eqnarray}
Thus the $s$-wave SC-NLSM system with a vortex line along the $z$-direction indeed has the inversion symmetry and the mirror symmetry. Similar results can be obtained for the $p$-wave SC-NLSM system.

In presence of a vortex line in the superconducting state, we introduce the following transformation:
\begin{eqnarray}
	\gamma_{{\bf \hat{r}},\sigma,1}&&=\gamma^{\dagger}_{{\bf \hat{r}} ,\sigma,1}=\frac{1}{\sqrt{2}}(c^{\dagger}_{{\bf \hat{r}},\sigma}+c_{{\bf \hat{r}},\sigma}) \\
	\gamma_{{\bf \hat{r}},\sigma,2}&&=\gamma^{\dagger}_{{\bf \hat{r}},\sigma,2}=\frac{i}{\sqrt{2}}(c^{\dagger}_{{\bf \hat{r}},\sigma}-c_{{\bf \hat{r}},\sigma}),	
\end{eqnarray}
Where $\gamma_{{\bf \hat{r}},\sigma,1/2}$ represent Majorana operator. Substituting Eqs.(S15) and (S16) into Eq. (S9), we have
\begin{eqnarray}
	H^{p}(k_{z}=0/\pi)=i\gamma_{k_{z}=0/\pi}H_{M}^{p}(k_{z}=0/\pi)\gamma^{\dagger}_{k_{z}=0/\pi}. \nonumber \\
\end{eqnarray}
Here $\gamma_{k_{z}=0/\pi}=\left[ \gamma_{1a1},\gamma_{1a2} ,\gamma_{1b1},\gamma_{1b2},\dots, \gamma_{{\bf \hat{r}}a1},\gamma_{{\bf \hat{r}}a2},\gamma_{{\bf \hat{r}}b1},\gamma_{{\bf \hat{r}}b2} \right]. $ Rewriting the $p$-wave SC-NLSM Hamiltonian in the Majorana representation($k_{z}=0/\pi$), we have
\begin{eqnarray}
	H_M^{p}(k_{z})=&&\sum_{{\bf \hat{r}, \sigma}}[-2i\mu\gamma_{{\bf \hat{r}},\sigma,1}\sigma_{0}\gamma_{{\bf \hat{r}},\sigma,2}+2(m-6+2\cos(kz))i\gamma_{{\bf \hat{r}},\sigma,1}\sigma_{x}\gamma_{{\bf \hat{r}},\sigma,2}+2t_{x}i\gamma_{{\bf \hat{r}},\sigma,1}\sigma_{x}\gamma_{{\bf \hat{r}+\hat{x}},\sigma,2} \nonumber \\
	&&-2t_{x}i\gamma_{{\bf \hat{r}},\sigma,2}\sigma_{x}\gamma_{{\bf \hat{r}+\hat{x}},\sigma,1}+2t_{y}i\gamma_{{\bf \hat{r}},\sigma,1}\sigma_{x}\gamma_{{\bf \hat{r}+\hat{y}},\sigma,2}-2t_{y}i\gamma_{{\bf \hat{r}},\sigma,2}\sigma_{x}\gamma_{{\bf \hat{r}+\hat{y}},\sigma,1}\nonumber \\
	&&-2{\mathrm{Re}}\Delta_{p}({\bf \hat{r}+\frac{\hat{x}}{2}})i(\gamma_{{\bf \hat{r}},\sigma,1}\sigma_{0}\gamma_{{bf \hat{r}+\hat{x}},\sigma,1}-\gamma_{{\bf \hat{r}},\sigma,2}\sigma_{0}\gamma_{{\bf \hat{r}+\hat{x}},\sigma,2}) \nonumber \\
	&&-2{\mathrm{Im}}\Delta_{p}({\bf \hat{r}+\dfrac{\hat{x}}{2}})i(\gamma_{{\bf \hat{r}},\sigma,1}\sigma_{0}\gamma_{{\bf \hat{r}+\hat{x}},\sigma,2}+\gamma_{{\bf \hat{r}},\sigma,2}\sigma_{0}\gamma_{{\bf \hat{r}+\hat{x}},\sigma,1}) \nonumber \\
	&&+2{\mathrm{Im}}\Delta_{p}({\bf \hat{r}+\frac{\hat{y}}{2}})i(\gamma_{{\bf \hat{r}},\sigma,1}\sigma_{0}\gamma_{{\bf \hat{r}+\hat{y}},\sigma,1}-\gamma_{{\bf \hat{r}},\sigma,2}\sigma_{0}\gamma_{{\bf \hat{r}+\hat{y}},\sigma,2}) \nonumber \\
	&&-2{\mathrm{Re}}\Delta_{p}({\bf \hat{r}+\dfrac{\hat{y}}{2}})i(\gamma_{{\bf \hat{r}},\sigma,1}\sigma_{0}\gamma_{{\bf \hat{r}+\hat{y}},\sigma,2}+\gamma_{{\bf \hat{r}},\sigma,2}\sigma_{0}\gamma_{{\bf \hat{r}+\hat{y}},\sigma,1})]. \nonumber \\
\end{eqnarray}

The topology of the vortex line is described by the ${\mathrm Z_{2}}(d=1)$ invariant, expressed as~\cite{Fu_2007a,Kitaev_2001,Budich_2013},
\begin{eqnarray}
	{\mathrm v}_{p}={\mathrm{sgn}}\left\lbrace {\mathrm{Pf}}\left[ H^{p}_{M}(k_{z}=\pi) \right]  \right\rbrace {\mathrm{sgn}}\left\lbrace {\mathrm{Pf}}\left[ H^{p}_{M}(k_{z}=0) \right]  \right\rbrace. \nonumber \\
\end{eqnarray}
The Pfaffian number is abbreviated as ${\mathrm{Pf}}$. It is calculated based on the algorithm developed by Ref.~\cite{Wimmer_2012}.

\subsection{The size of flat band regions in the spinless $s$-wave NLSM superconductor}
The Hamiltonian of the spinless $s$-wave SC-NLSM in the main text is expressed as,
\begin{eqnarray}
	H^{s}= \sum_{\bf k}[M({\bf k})\sigma_x\otimes\tau_{z}+\lambda_{z}\sigma_{z}\otimes\tau_{0}-\mu\sigma_{0}\otimes\tau_{z}+\Delta_{s}\sigma_{y}\otimes\tau_{y}].\nonumber\\
\end{eqnarray}
For simplification, we set $\mu=0$. Diagonalizing $H^{s}$, we have\\
\begin{eqnarray}
	E=\pm \sqrt{M^{2}({\bf k})\pm2M({\bf k})\Delta_{s}+\lambda_{z}^{2}+\Delta_{s}^{2}}
\end{eqnarray}
At $E=0$, it generates two rings with radios being $R_{1,2}=M({\bf k})\pm\Delta_{s}$. At the system surface,
the flat bands appear between two nodal rings, as is seen in Figs.~2(c) and 2(e) of the main text.
The size of the flat band regions can be calculated as $\delta=R_{1}-R_{2}=2\Delta_{s}$.
Therefore, the size of flat band regions with a nonzero chemical potential is in proportion to the $s$-wave superconducting gap. As a result, at the system surface, the size of the flat band region will increase
monotonously when the gap magnitude increases. This explain qualitatively that in the spinless $s$-wave SC-NLSM
with a vortex line, the zero energy LDOS reaches the maximum value at the system corners, as is seen in Fig.~4(e) of the main text.

\end{document}